\documentclass[12pt,preprint]{aastex}

\usepackage{multicol}

\shorttitle{Accelerated Evolution in Compact Groups}
\shortauthors{Walker et al.}

\begin{document}

\title{Mid-Infrared Evidence for Accelerated Evolution in Compact Group Galaxies}

\author{Lisa May Walker and Kelsey E. Johnson}
\affil{Department of Astronomy, University of Virginia,
    Charlottesville, VA 22904}
\author{Sarah C. Gallagher}
\affil{Department of Physics and Astronomy, University of Western Ontario,
    London, ON N6A 3K7 Canada}
\author{John E. Hibbard}
\affil{National Radio Astronomy Observatory,
    Charlottesville, VA 22903}
\author{Ann E. Hornschemeier and Panayiotis Tzanavaris}
\affil{Laboratory for X-Ray Astrophysics, NASA Goddard Space Flight Center,
    Greenbelt, MD 20771}
\author{Jane C. Charlton}
\affil{Department of Astronomy and Astrophysics, Pennsylvania State University,
    University Park, PA 16802}
\and
\author{Thomas H. Jarrett}
\affil{\textit{Spitzer} Science Center, California Institute of Technology,
    Pasadena, CA 91125}

\begin{abstract}
Compact galaxy groups are at the extremes of the group environment, with high number densities and low velocity dispersions that likely affect member galaxy evolution. To explore the impact of this environment in detail, we examine the distribution in the mid-infrared (MIR) $3.6-8.0\;{\rm \mu m}$ colorspace of 42 galaxies from 12 Hickson compact groups in comparison with several control samples, including the LVL+SINGS galaxies, interacting galaxies, and galaxies from the Coma Cluster. We find that the HCG galaxies are strongly bimodal, with statistically significant evidence for a gap in their distribution. In contrast, none of the other samples show such a marked gap, and only galaxies in the Coma infall region have a distribution that is statistically consistent with the HCGs in this parameter space. To further investigate the cause of the HCG gap, we compare the galaxy morphologies of the HCG and LVL+SINGS galaxies, and also probe the specific star formation rate (SSFR) of the HCG galaxies. While galaxy morphology in HCG galaxies is strongly linked to position with MIR colorspace, the more fundamental property appears to be the SSFR, or star formation rate normalized by stellar mass. We conclude that the unusual MIR color distribution of HCG galaxies is a direct product of their environment, which is most similar to that of the Coma infall region. In both cases, galaxy densities are high, but gas has not been fully processed or stripped. We speculate that the compact group environment fosters accelerated evolution of galaxies from star-forming and neutral gas-rich to quiescent and neutral gas-poor, leaving few members in the MIR gap at any time.
\end{abstract}


\section{INTRODUCTION}
Galaxies form and evolve in a wide range of environments, from sparse field to densely populated groups and clusters. The most extreme densities are encountered in the cores of rich clusters, which are relatively rare in number, and in the more numerous compact groups \citep{hickson82}. The Hickson Compact Groups (HCGs), owing to their high number density coupled with low velocity dispersions, undergo frequent tidal interactions, distortions and mergers between group members \citep{hickson92, mendes94}. HCGs are unique laboratories to study extreme galaxy evolution in the local universe and may also serve as analogs to galaxy formation and evolution in the early universe when population densities were much higher than what is observed in the field today.

A challenge of studying HCGs is moving from a morphology-based to a quantitatively based description of the star-formation and evolution within the extreme population-density environment. \citet{arp} find that the mass-normalized star formation rates (SFR) in interacting galaxies are approximately twice that in normal spirals, lending support to the idea that interactions induce star formation. However, \citet{rosa07} and others find evidence of a mechanism that quenches the star formation in compact groups. Specifically, they find that the stellar populations in elliptical galaxies in HCGs are more metal-poor and older than their counterparts in the field. Thus, although compact groups might be thought to be an ideal site for merger-induced star formation, this does not always seem to be the case.

In order to assess the impact of the compact group environment on star formation, \citet{hcgs} looked at the {\it Spitzer} IRAC ($3.6-8.0\;{\rm \mu m}$) colorspace distribution of HCGs and found that the mid-infrared (MIR) colors of galaxies in \ion{H}{1} gas-rich HCGs are dominated by star formation, while the MIR colors of galaxies in \ion{H}{1} gas-poor HCGs are dominated primarily by stellar photospheric emission, or are MIR-passive. From this, they infer an evolutionary sequence in which the gas in \ion{H}{1}-rich groups is consumed, expelled, or ionized. In a complementary study of loose groups at higher redshift (z$\sim$0.4), \citet{wilman08} see a bimodality in {\it k}-corrected $\left[f\left(8.0\right)/f\left(3.6\right)\right]$ colors, and a deficit of infrared activity when compared to field galaxies, similar to local \ion{H}{1}-poor compact groups. They note that the fraction of infrared excess galaxies, $f\left(IRE\right)$ decreases with galaxy stellar mass $M_*$, but within their group sample they see a deficit in $f\left(IRE\right)$ at all masses, and state that this trend can be explained if suppression of $M_* > 10^{11} M_\sun$ galaxies occurs primarily in the group environment.

In {\it Spitzer} IRAC color-color plots of HCG galaxies, \citet{hcgs} noted a ``gap'' in the distribution of galaxies between those dominated by stellar light and galaxies that are actively star-forming. This gap did not appear to be present in the initial comparison sample from the \textit{Spitzer} First Look Survey (FLS). This comparison led them to speculate that the gap may be due to rapid evolution caused by the unique dynamical influences present in HCGs. However, the range in redshift of the FLS galaxies, which shifts polycyclic aromatic hydrocarbon (PAH) features in and out of the observing bands, renders the plot ambiguous. Nonetheless, \citet{gallagher08}, using a related mid-IR diagnostic, confirmed the discrepancy by comparing HCG galaxies to the local SINGs sample \citep{sings}, which like the FLS sample exhibits no gap in mid-IR color space. \citet{tzanavaris10} determined SFRs and specific SFRs (SSFRs) for this same HCG sample using {\it Swift} UV and {\it Spitzer} 24 ${\rm \mu m}$ data and found a gap in SSFRs between $3.2\times10^{-11}\;{\rm yr^{-1}}$ and $1.2\times10^{-10}\;{\rm yr^{-1}}$. This gap did not exist in their comparison sample comprised of SINGS non-interacting and isolated galaxies. The fact that the HCG sample is discrepant with both the FLS and SINGS samples led them to conclude that the local environment in HCGs strongly influences the member galaxies.

Galaxies in compact groups likely evolve differently than galaxies in other environments. Galaxies in clusters also experience frequent interactions, but the interaction timescale is shorter due to the large velocity dispersions, and they reside in a smoother gravitational potential, causing less torquing on the gas within galaxies. On the other hand, galaxies in loose groups experience less frequent interactions on average than those in compact groups. Giant field galaxies experience fewer major interactions, and frequently dominate their local gravitational potential. Comparing galaxies in high density environments (e.g. clusters, compact groups) with galaxies in medium density environments (e.g. interacting pairs of galaxies) and low density environments (e.g. individual field galaxies) may allow greater understanding of how galaxy evolution is affected by the local density.

In an effort to learn more about the nature of the gap found by \citet{hcgs}, we have examined the MIR colors of other samples of galaxies in a variety of environments. The previous comparison samples, FLS \citep{hcgs} and SINGS \citep{gallagher08, tzanavaris10}  were both limited in scope and are known to suffer from issues with the samples that could have led to their discrepancy with the HCG sample. The goal of this paper is to expand the comparisons to a range of other samples with different properties in order to better assess the influence of environment on MIR colorspace distribution. We use the Kolmogorov-Smirnov test to compare the MIR colors of the galaxies in these samples with those of the HCG galaxies to investigate to what extent the gap seen by \citet{hcgs} is ubiquitous, or unique to the compact group environment.

\section{DATA/SAMPLES}
\begin{figure}
  \plotone{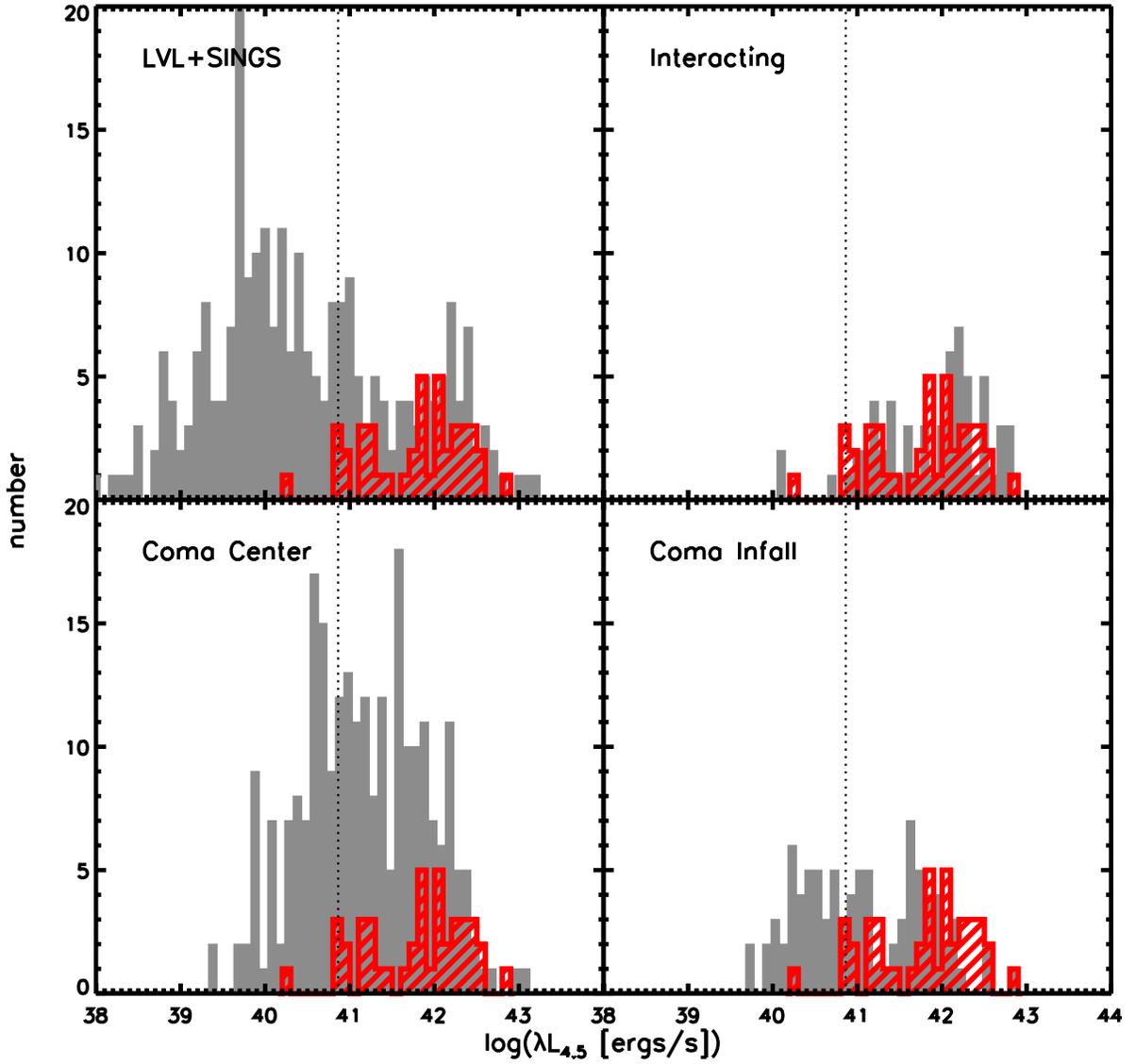}
\caption{Histograms of $\lambda L_{4.5}$ for each sample, with the HCG galaxies overlaid with the red striped histogram. The histograms  yield similar results regardless of whether $\lambda L_{3.6}$ or $\lambda L_{4.5}$ is used. The dotted vertical line indicates the minimum luminosity for comparison and statistical analysis between samples.\label{lumhist}}
\end{figure}

\begin{figure}
  \plotone{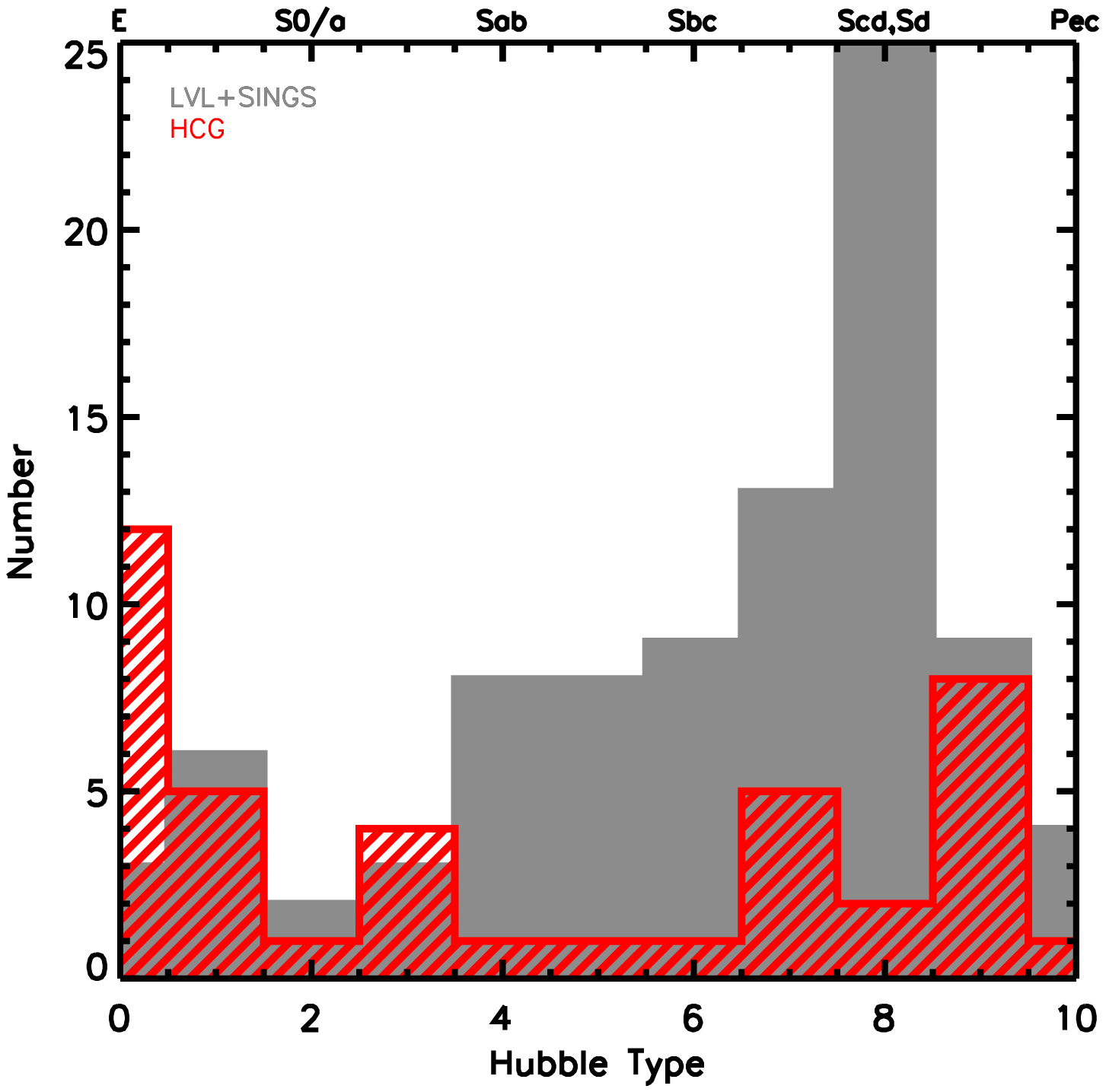}
\caption{Histograms of Hubble type \citep[following][]{haynes84} for the HCG galaxies (striped) and LVL+SINGS galaxies (solid).\label{hubblehist}}
\end{figure}

\subsection{HCG Galaxies}
\label{hcgobs}
The HCG dataset, taken from \citet{hcgs}, comprises 42 galaxies from 12 groups. The groups contain varying amounts of \ion{H}{1}, and span the three classification categories discussed in \citet{hcgs}. The most \ion{H}{1} gas-rich groups with $\log{\left( M_{\mathrm{H \, I}} \right)}/\log{\left( M_{\mathrm{dyn}} \right)} \ge 0.9$ are classified as type I , the \ion{H}{1} gas-poor groups with $\log{\left( M_{\mathrm{H \, I}} \right)}/\log{\left( M_{\mathrm{dyn}} \right)} < 0.8$ as type III, while type II contains the intermediate groups. It is also important to note that these categories reflect the {\it group} \ion{H}{1}, which is not necessarily simply the sum of the \ion{H}{1} content of individual galaxies as this classification also includes gas in the IGM. Table \ref{hcgsample} gives background data on these groups. NED has been searched for known AGN, and four Seyfert IIs were identified: HCG 16b, 16d, 61a, and 90a. However, we expect emission from the AGN in these cases should be minimal compared to the integrated light of the galaxy, as found by \citet{gallagher08}.

One question brought up in discussions of compact groups (CGs), especially HCGs, is the effect of selection biases. \citet{mamon94} claims that the properties of CGs are a function of the algorithm used to find them. \citet{ribeiro98} discusses Hickson's original selection criteria and how they influence the catalog of HCGs, such as ignoring the faint end of the luminosity function, selecting only groups with ``significant surface density enhancement over the field,'' and appearing to be well isolated, taking them out of context with their surroundings. Since Hickson's original catalog, several other catalogs of CGs have been published, some based on sky position and photometry like Hickson's, some utilizing redshift data. With the flood of data from SDSS, several catalogs of CGs have been generated, allowing selection effects to be qualitatively studied. \citet{lee04} selected CGs from the SDSS commissioning data using criteria slightly modified from Hickson's; they instituted an upper limit on number of group members, changed the isolation criteria to greater than three angular diameters, rather than greater than or equal to, and required higher surface brightness. They compared their catalog with six existing catalogs of CGs, including the HCGs, which they acknowledge as ``the benchmark for all CG catalogs.'' For each CG catalog, they looked at the mean and median of group members, surface brightness, redshift, angular diameter, and linear diameter, and the HCGs tend to fall in the middle of each distribution \citep[Table 3 in][]{lee04}. Thus, it appears that any selection biases that affect the HCGs are either more strongly present in other catalogs, or inherent properties of CGs as a class. In the latter case, this can be interpreted as reflecting the role of the CG environment on the evolution of the constituent galaxies.

\input{hcgsample}

\subsection{Comparison Samples}
\label{compobs}
Several comparison samples were selected from previous \textit{Spitzer} IRAC studies. We have endeavored to include surveys that target a range of galaxy densities in order to begin to differentiate the role of environment on the observed MIR properties. These comparison samples include the Local Volume Legacy Survey (LVL) + \textit{Spitzer} Infrared Nearby Galaxies Survey (SINGS) galaxies \citep{lvl, sings}, interacting galaxies \citep{arp}, and galaxies in the Coma cluster \citep{coma}. For both the HCG galaxies and the comparison samples, we only included galaxies at $z < 0.035$, in order to ensure that the PAH features have not been redshifted into or out of their rest-frame IRAC channels.

The LVL data, discussed in \citet{lvl}, consists of 258 galaxies within 11 Mpc. Galaxies with undefined flux values or only upper limits on the photometry were excluded, leaving 211 galaxies in the sample. The SINGS data, as presented and described in \citet{sings}, consist of 71 galaxies. Four of the galaxies in the original dataset (M81 Dwarf A, NGC 3034, Holmberg IX, and DDO 154) were excluded because only upper or lower limits on the photometry were provided. It is important to note that the SINGS sample was chosen to be diverse, which will affect its distribution in colorspace. In addition, we combined the LVL and SINGS galaxies to create a control sample, referred to as LVL+SINGS. The 35 galaxies from the Spitzer Spirals, Bridges, and Tails Interacting Galaxy Survey (hereafter referred to as the interacting sample) are comprised of otherwise relatively isolated binary galaxy systems, whose members are tidally disturbed \citep{arp}. This sample was biased towards galaxies with prominent signs of interaction, thereby selecting against elliptical galaxies. The Coma sample, discussed in \citet{coma}, is comprised of galaxies from two fields. The first field is located in the center of the cluster, where the galaxy density is very high. The second field is the infall region, located near 0.4 virial radii at the secondary X-ray peak, where the galaxy density is still higher than field density \citep{coma}. Two galaxies from the center of the Coma cluster have been removed from the sample due to uncertain apertures, caused by proximity to one of the central elliptical galaxies. Our HCG sample has only one galaxy with luminosity below $\log{\left(L_{4.5}\;[\rm{erg/s}]\right)} = 40.9$. With Spitzer, we mapped the entire extent of each group to a 4.5 micron sensitivity better than this limit \citep{hcgs}, thus our HCG database should be complete to this level. To be conservative, to compare the HCG galaxies to similar galaxy populations from the other samples we only consider galaxies with higher luminosities. The luminosity distributions of the samples are shown in Figure \ref{lumhist}, with a vertical dotted line indicating the minimum luminosity required for inclusion. Figure \ref{hubblehist} shows the distribution of Hubble types in our HCG sample compared with the LVL+SINGS sample. As can be seen in this plot, the LVL+SINGS sample is deficient in Sa galaxies, with respect to other Hubble types. The LVL+SINGS sample is dominated by spirals later than Sa, while the HCG sample contains galaxies across the distribution.

\section{MID-INFRARED COLORSPACE}\label{colorspace}
\subsection{HCG Galaxies}
\begin{figure}
  \plotone{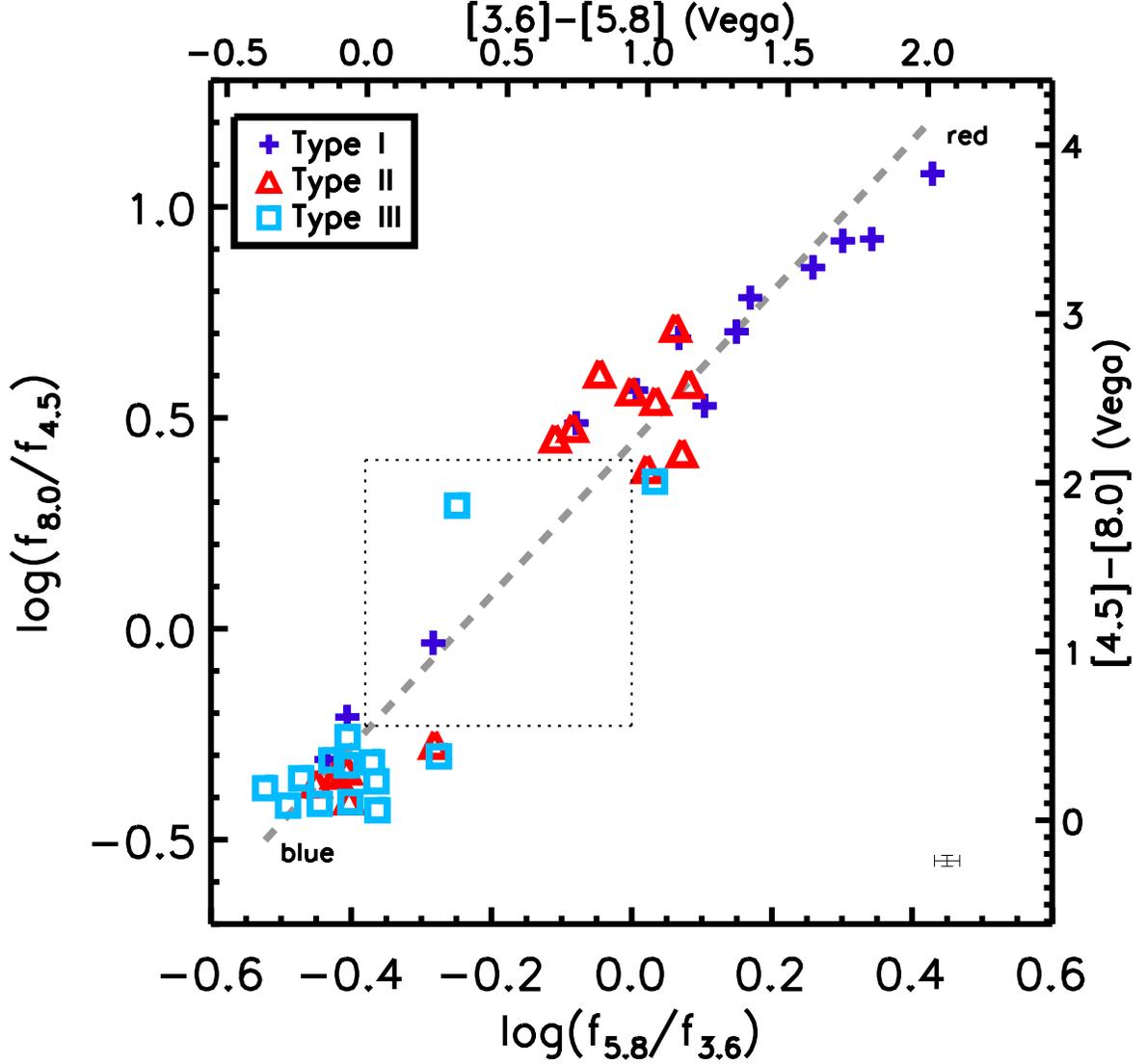}
\caption{Color-color plot of HCG galaxies. The indigo plus signs are galaxies from type I (\ion{H}{1} gas-rich) groups, the red triangles correspond to galaxies from type II groups, and the blue squares represent galaxies from type III (\ion{H}{1} gas-poor) groups. The error bars in the lower right of the plot indicate typical errors for the sample. The lower-left region of the plot contains galaxies whose light is dominated by stellar photospheric emission, while actively star-forming galaxies reside in the upper right. Between these is the ``gap'' noted by \citet{hcgs}, indicated by the dotted box. The dashed line is the linear fit to the data, used for coordinate rotation in \S\ref{ks}.\label{hcgcol}}
\end{figure}

The IR color-color plot for the HCG galaxies is shown in Figure \ref{hcgcol}. The galaxies are separated into three types based on the fractional \ion{H}{1} mass of the groups they belong to - the indigo plus signs represent type I (\ion{H}{1} gas-rich), the red triangles indicate type II, and the blue squares correspond to type III (\ion{H}{1} gas-poor). Galaxies in the lower-left portion of the plot have IR spectral energy distributions (SEDs) consistent with being dominated by stellar light. \ion{H}{1} gas-poor groups fall in this region of the diagram due to the higher percentage of E/S0 galaxies. Galaxies with ongoing star formation will tend to fall in the upper right with red MIR colors indicative of interstellar polycyclic aromatic hydrocarbon (PAH) emission and thermal emission from warm/hot dust. The majority of the galaxies in the active region of the plot come from both type I (\ion{H}{1}-rich) and type II groups, though there is a trend that galaxies from type II groups fall in the blue part of the active region, while the galaxies from type I groups show a larger range of colors throughout the region of the plot indicating activity. This trend can also be attributed to the percentage of E/S0 galaxies, as the type I groups contain very few E/S0 galaxies, while the type II groups have a varying percentage of E/S0 galaxies.

As \citet{hcgs} noted, it is apparent in Figure \ref{hcgcol} that the region between the galaxies dominated by stellar emission and galaxies with active star formation contains relatively few galaxies. In the following sections, we will explore the hypothesis that this gap in the MIR colors of HCG galaxies is due to rapid evolution through the stage during which galaxies have intermediate MIR colors, or in other words, the HCG environment is biased against galaxies with very modest amounts of star formation.

\subsection{Comparison Samples}
\begin{figure}
  \plotone{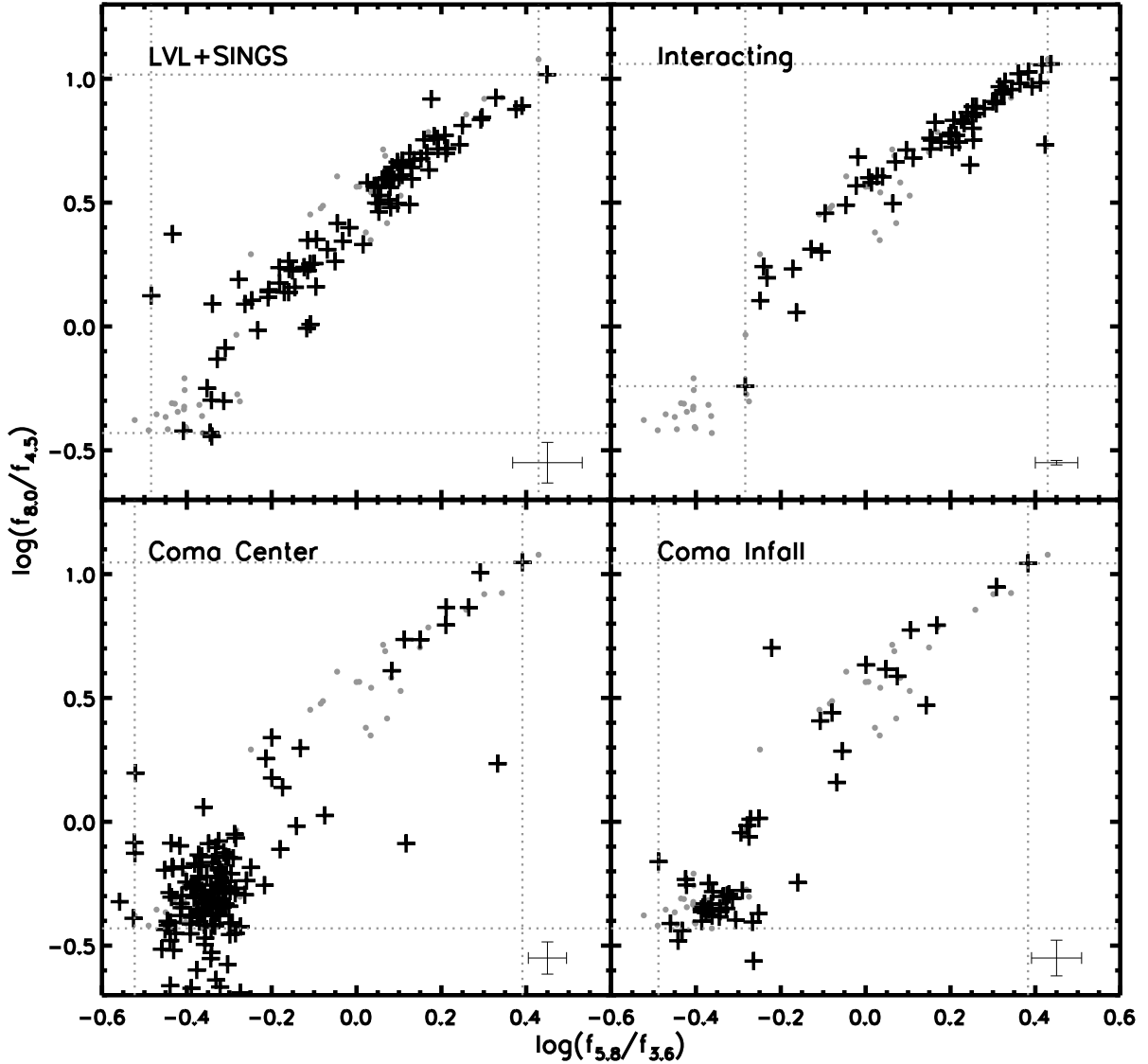}
\caption{Color-color plot of the comparison samples. The grey dots are the HCG galaxies, and the error bars in the bottom right indicate typical errors for each comparison sample. The dotted lines indicate the cropping values, discussed in \S\ref{ks}. The LVL+SINGS sample is distributed rather uniformly in colorspace. The interacting sample is located primarily in the active (upper right) region of the plot. The two Coma samples have a significant population in the stellar colors (lower left) region of colorspace; central Coma galaxies show no sign of a gap, while the infall Coma sample shows an underdensity of points in the same location as the HCG gap.\label{colcolplots}}
\end{figure}

The LVL+SINGS galaxies span the same region in MIR colorspace as the HCG galaxies, as shown on the top left in Figure \ref{colcolplots}. Given that the LVL is a volume-limited survey and SINGS galaxies were chosen to represent a range of physical properties and environments, it is not surprising that they are distributed relatively uniformly in colorspace rather than clustered in one region. Figure \ref{colcolplots} does not show any evidence of a gap in colorspace for the LVL+SINGS galaxies. A plot made with the culled SINGS sample of \citet{arp} (not shown) is similar, and also does not show a gap. The culled sample was created by removing galaxies with nearby companions from the SINGS sample.

The sample of interacting galaxies (top right plot of Figure \ref{colcolplots}) forms a very tight relationship in the star-forming region of colorspace, which is consistent with the fact that interactions frequently trigger star formation. There is only one point whose colors are consistent with being dominated by stellar populations, with little, if any, PAH emission and thermal emission from warm dust. The selection against elliptical galaxies mentioned in \S\ref{compobs} also likely contributes to the lack of points in the blue region of IRAC colorspace.

The majority of the galaxies in the center of the Coma cluster lie in the region of MIR colorspace corresponding to stellar colors, with a smaller number of the galaxies scattered throughout the redward side of the plot. This is consistent with the fact that galaxies in the cores of clusters tend to contain little to no \ion{H}{1} and star formation. Unlike the other samples we consider, the central Coma galaxies do not follow a tight relationship in colorspace but rather have a much more scattered distribution. There are two galaxies with unusual colors: red $\log(f_{5.8}/f_{3.6})$ and 
blue $\log(f_{8.0}/f_{4.5})$. These two galaxies do not appear anomalous in the IRAC data, and we were unable to find anything in the literature which might explain their odd location in IRAC colorspace. However, both of these galaxies are in very crowded fields, and that may be the cause of their odd colors. The central Coma galaxies are rather smoothly distributed, with no sign of a gap in their distribution.

Despite being from the same galaxy cluster, the distribution in colorspace of galaxies in the infall region of the Coma cluster is less concentrated in the region of stellar colors than the center of the Coma cluster. There are galaxies whose light is dominated by stellar emission, but the star-forming region for the infall sample is relatively more populated than the central Coma region. The color-color plot of galaxies in the infall region of the Coma cluster reveals an underdensity of points in the same location as the gap we see in the HCG galaxies. While it does not seem to be as pronounced, it may still carry some significance.

\section{STATISTICAL TESTS}\label{ks}
In order to assess whether the gap in the distribution of HCG galaxies seen in Figure \ref{hcgcol} is significant, we perform a statistical analysis using the Kolmogorov-Smirnov (KS) test. The IDL routines \texttt{ksone} and \texttt{kstwo} were used to perform the KS test, which first required rotating the data. The single-distribution version of the test allows comparison of each sample to a model of a uniform distribution, and determines whether it is an accurate description of the data. The two-distribution version of the test compared two samples to determine whether they could have been drawn from the same distribution. Notably, it is only possible to conclusively reject the null hypothesis using this method; the KS test cannot confirm the hypothesis.

We performed the two-distribution test in two ways: 1) comparing only the region in colorspace that is populated by both the HCG sample and the comparison sample (indicated by the dotted lines in Figure \ref{colcolplots}), from which we cannot draw any conclusions about the distributions outside of this range; and 2) comparing the entire region of colorspace populated by either the HCG sample or the comparison sample. We fit a line to the HCG distribution in colorspace, which became the new x-axis in rotated colorspace, $\Delta C_{MIR}$.

It is important to note that a ``uniform'' distribution does not necessarily mean a ``normal'' or ``expected'' distribution. A uniform distribution of galaxies in colorspace would be a sample whose galaxies fall evenly along a trend in colorspace. However, a normal distribution for old elliptical galaxies would show the galaxies preferentially in a clump in the bottom-left, stellar colors region. In a normal distribution of active (star-forming) galaxies, all the galaxies would fall linearly in the upper-right, active region. What we wish to determine is which types of environments give rise to a gap.

\subsection{Single-Distribution KS Test}
\input{ksone}
For the single-distribution test, the cumulative distribution function (CDF) of the rotated data was compared with the CDF of a model representing a uniform distribution along the x-axis over the same range as the galaxies. The resulting test yields two values: $D$, which is the maximum deviation of the data from the model, and $\alpha$, which gives the significance level with which it is possible to reject the null hypothesis that the model matches the data.

\subsubsection{HCG Galaxies}
\begin{figure}
  \plotone{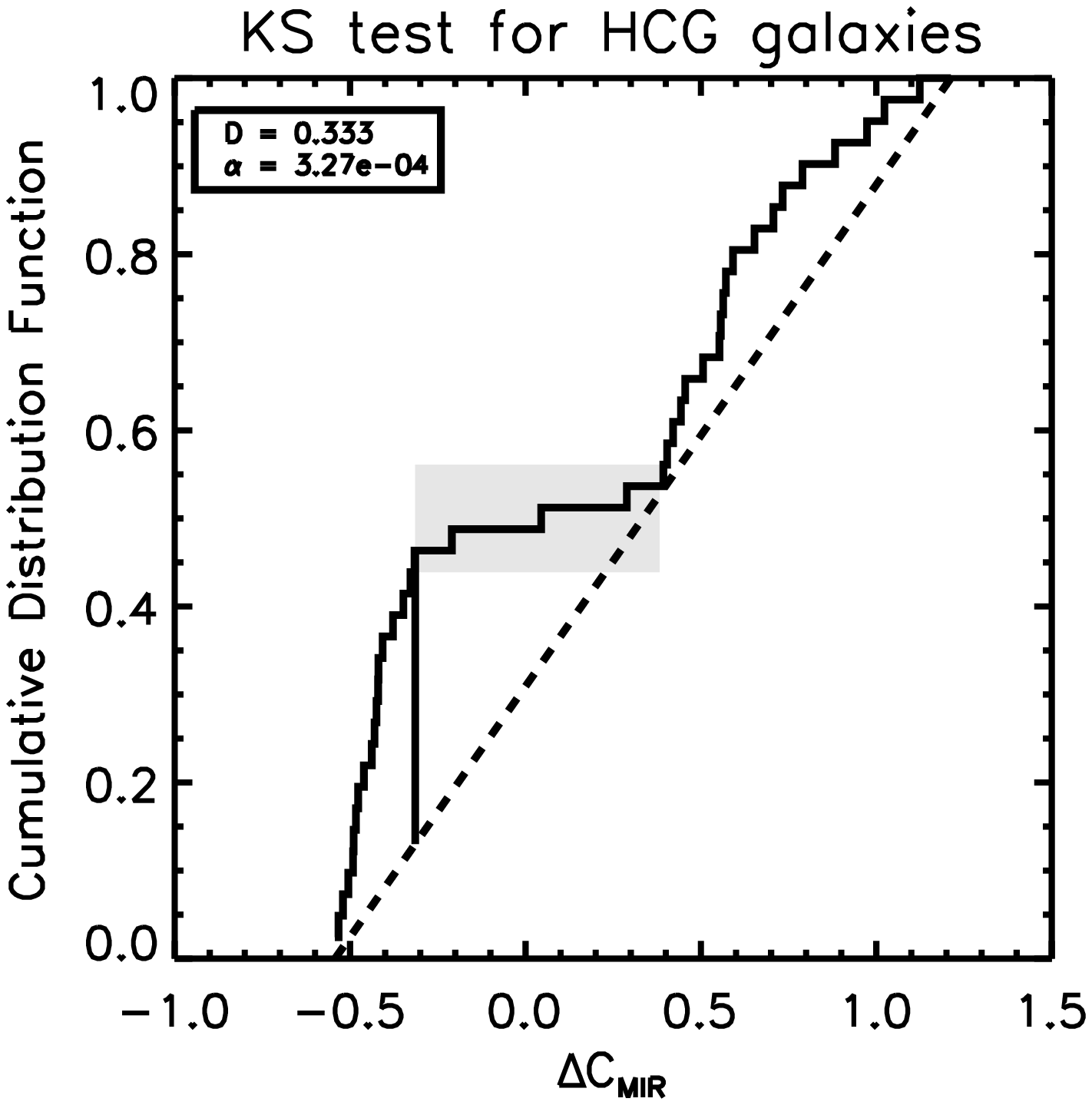}
\caption{KS test for the rotated HCG distribution against a model of uniform distribution. The maximum vertical distance between the CDF of the HCG galaxies (solid line) and that expected for a uniform distribution (dashed line) is $D$, indicated by the vertical line. The large value of $D$ indicates that the probability that the HCGs are drawn from a uniform distribution is very low, $0.03\%$. The nearly flat portion of the CDF highlighted in grey qualitatively reveals the gap.\label{hcgks}}
\end{figure}

Figure \ref{hcgks} shows the single-distribution KS test for the HCG galaxies. It clearly illustrates the gap apparent in Figure \ref{hcgcol}, manifested as the nearly horizontal portion of the CDF over $-0.31 < \Delta C_{MIR} < 0.38$ indicated by the grey rectangle. The maximum deviation of the sample from the model occurs at the beginning of the gap and is due to the pile-up of galaxies dominated by stellar light. The value of $\alpha$ returned by the test means that it is significant to reject the hypothesis that the HCG galaxies come from a uniform distribution at the 99.97\% confidence level.

\subsubsection{Comparison Samples}
\begin{figure}
  \plotone{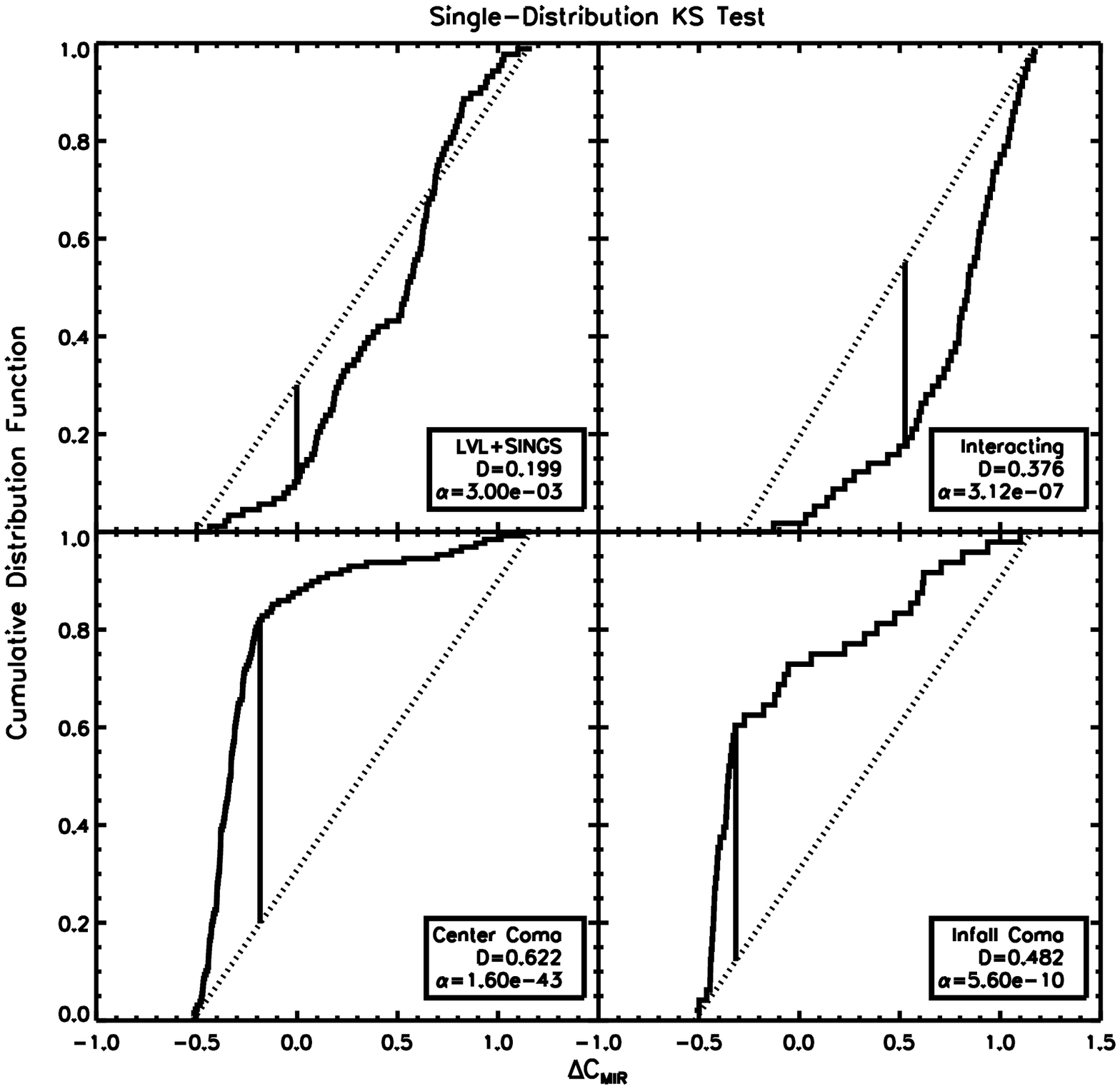}
\caption{KS test for the rotated comparison samples against a model of uniform distribution for the cropped samples, with D and $\alpha$ as defined in Figure \ref{hcgks}. The LVL+SINGS sample is mildly inconsistent. The interacting sample is concave because the galaxies tend to be gas-rich and MIR-bright, and therefore the distribution is weighted towards higher values of $\Delta C_{MIR}$, while the two Coma samples are very convex due to the large fraction of galaxies with stellar colors in the samples.\label{allks1lum}}
\end{figure}

As Table \ref{ksone} and Figure \ref{allks1lum} show, the LVL+SINGS sample is mildly inconsistent with being drawn from a uniform distribution over the color range it covers. The CDF of the interacting sample is concave, indicating that the galaxies are dust-rich and likely gas-rich, forming stars. The CDF of center Coma galaxies, and to a lesser extent the galaxies in the infall region of Coma, is very convex. This is due to the relative overabundance of galaxies with stellar colors, caused by the lack of cold gas and therefore star formation present in these galaxies. Thus the fact that the infall Coma region CDF is less convex than the center Coma region may indicate that galaxies in this region have not undergone as much processing. From the single-distribution test, it is clear that sample CDFs are affected by environment, as the HCG and Coma samples (i.e. the dense systems) have different CDFs from LVL+SINGS (i.e. the ``field'' sample).

\subsection{Two-Distribution KS Test}\label{ks2}
\input{kstwo}
The two-distribution KS test compared the CDF of each sample against the CDF of the HCG sample. This also calculated $D$ and $\alpha$, with the same meaning as the single-distribution test except that $D$ is now the maximum deviation between the two samples.

\begin{figure}
  \plotone{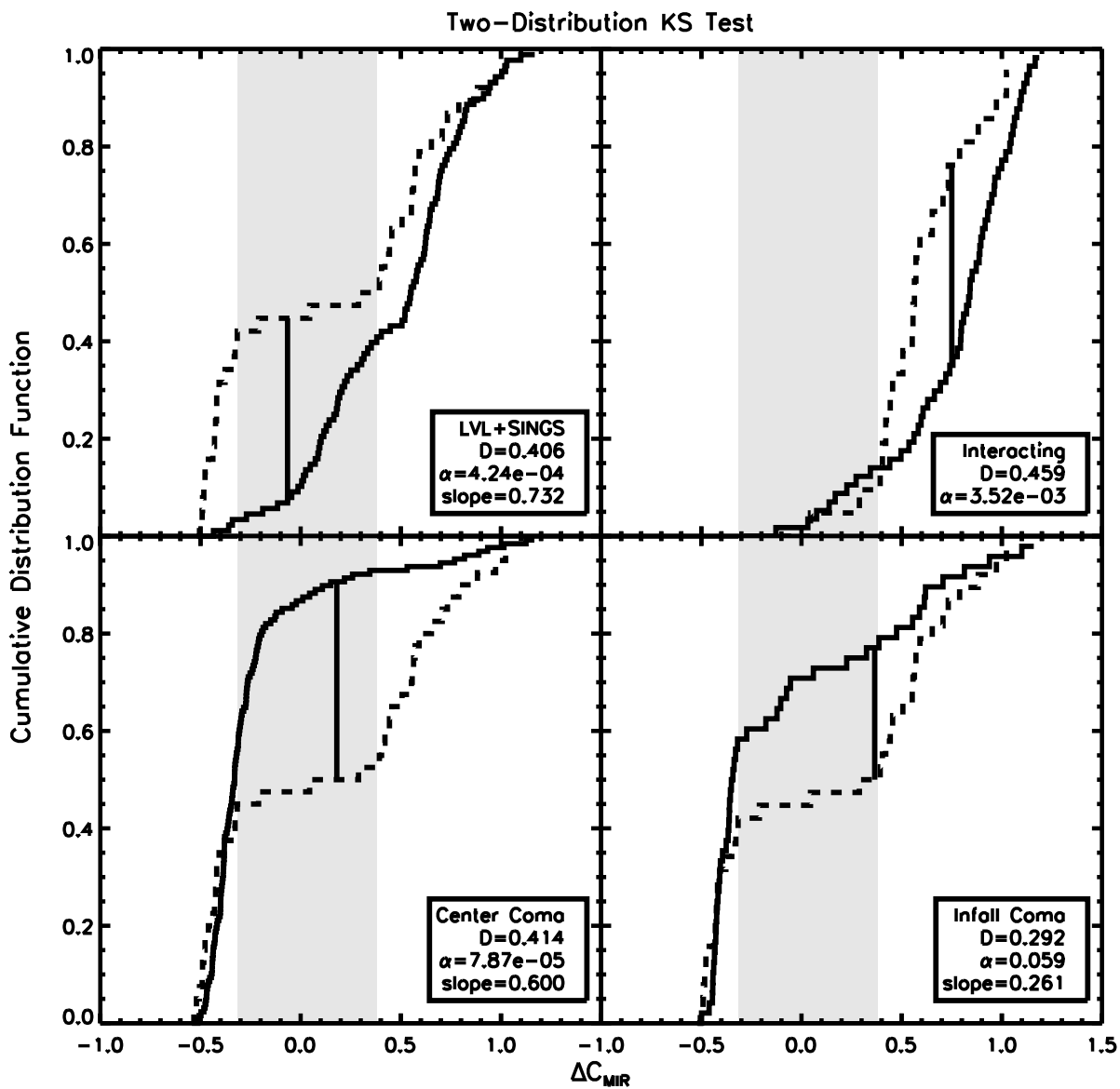}
\caption{KS test for the rotated comparison samples against the HCG galaxies for the cropped samples. From the large values of D and small values of $\alpha$, as defined in Figure \ref{hcgks}, the LVL+SINGS, interacting, and Coma center samples are clearly inconsistent with the HCG sample. The Coma infall sample has a high enough $\alpha$ that we cannot reject the hypothesis that it is consistent with the HCGs. The slope is a linear fit to the cumulative distribution function over the gap region.\label{allks2lum}}
\end{figure}

The results of the two-distribution KS test are given in Table \ref{kstwo} and plotted in Figure \ref{allks2lum}. We deemed a sample inconsistent with the HCG sample if its probability of being consistent with the HCGs was less than 1\%. It is apparent that the distribution of galaxies in the center of Coma is drastically different from the distribution of HCGs, and the distributions for the LVL+SINGS galaxies and interacting galaxies are mildly inconsistent with being drawn from the same distribution as the HCGs. The $\alpha$ value for the infall Coma region sample is high enough that we cannot reject the hypothesis that it is drawn from the same parent distribution. Thus the infall Coma region is most like the HCGs, as it has the smallest $D$ and largest $\alpha$.

\subsection{Gap Region}
\input{gapslope}
In order to discover the depth of the gap seen in HCG colorspace, we devised a test to quantitatively discern whether any of the comparison samples exhibit a similar gap. We show the gap region by the grey box in Figure \ref{hcgks}. This region was defined by the two HCG galaxies bounding the gap, at $\Delta C_{MIR} = -0.31$ and $0.38$. Since the gap represents a deficit of galaxies over this color range, its signature is a flattening in the CDFs. Therefore as another quantitative measure of the gap we performed a linear fit to the CDF between the color boundaries to obtain the slope; shallower slopes indicate samples with more pronounced gaps. The results are given in Table \ref{gapslope}, which shows that the HCG sample has the most pronounced gap of any of the samples. We have excluded the interacting sample as it begins mid-gap (see Figure \ref{colcolplots}), so a slope would be undefined. The remaining three samples all contain galaxies in the region of stellar colors and thus cover the entire gap region. The slope of the CDF of the infall Coma region over the gap is fairly shallow, possibly indicating the existence of a less-defined gap than the HCG galaxies.

\section{COLOR-MAGNITUDE DIAGRAMS}\label{cmds}
\begin{figure}
  \plotone{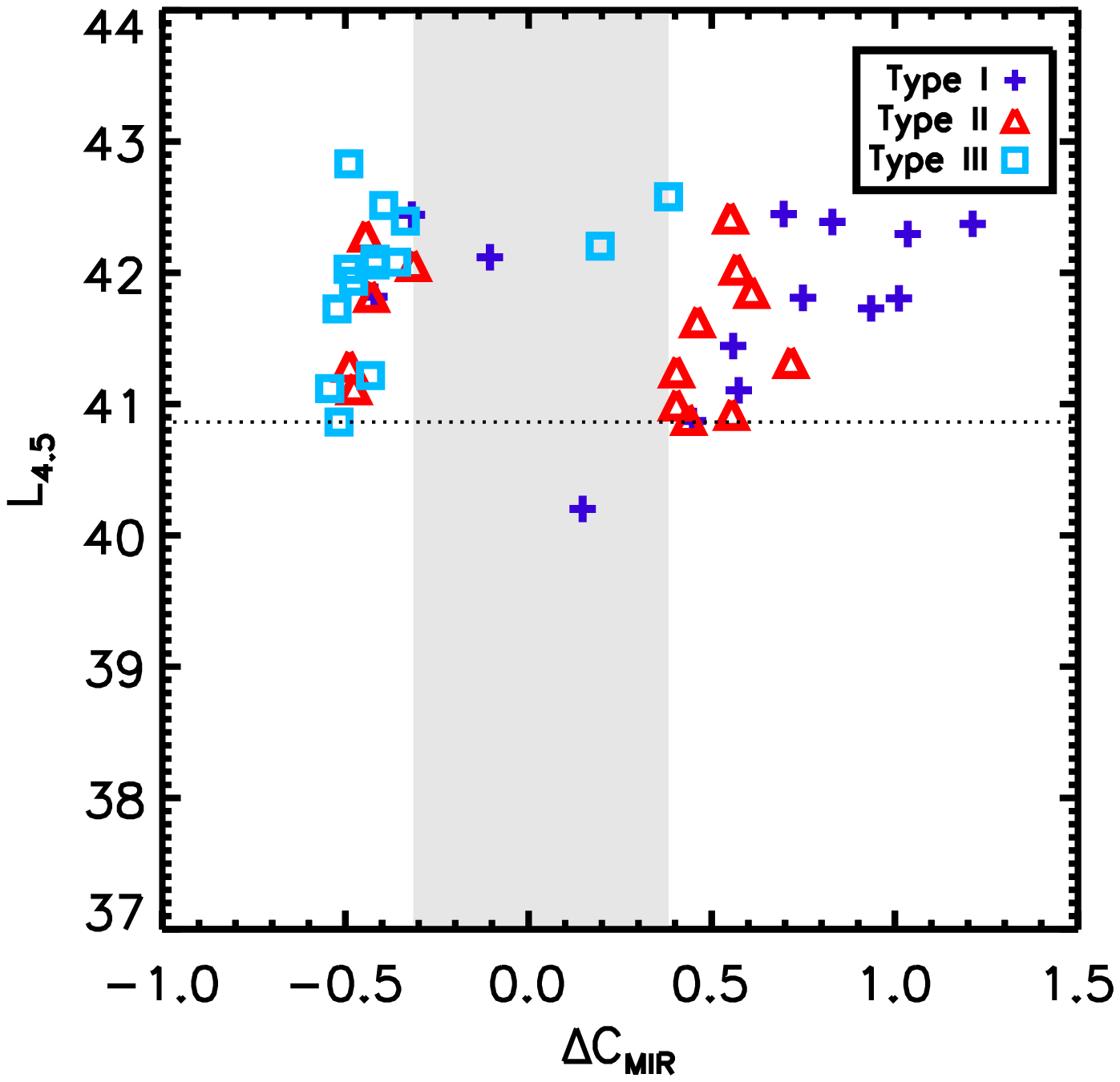}
\caption{CMD for the HCG galaxies. The dotted line indicates the minimum luminosity required for inclusion. The gap is still apparent, indicated by the grey stripe. Notably, there is no evidence for a color-luminosity correlation.\label{hcgcmd}}
\end{figure}

Optical color-magnitude diagrams have proven useful for understanding how galaxy luminosity (as a proxy for stellar mass) and color (as an indicator of the level of current star formation) are related. Galaxies are not evenly distributed in this parameter space, but are found in distinct regions: the so-called red sequence, blue cloud, and green valley \citep{hogg04}, discussed in \S\ref{optcmds}.  We are interested in investigating whether a color-magnitude diagram using only mid-IR information is similarly helpful in elucidating galaxy properties.

We created a plot analogous to a color-magnitude diagram using the data discussed in \S\ref{ks}. A comparison to the optical red sequence will be discussed in \S\ref{discussion}. Figure \ref{hcgcmd} shows the rotated color-luminosity diagram (hereafter referred to as CMD) for the HCGs, with the gap region again highlighted in grey. The color distribution and gap seen in Figure \ref{hcgcol} are clearly apparent in Figure \ref{hcgcmd}. Within the HCG sample, a galaxy's luminosity does not appear to depend on its color. Note that even if we relax the luminosity cut ($\log{\left(L_{4.5}\;[\rm{erg/s}]\right)} \ge 40.9$), the gap is still quite obvious. Galaxies to the left of the gap have blue MIR colors, indicating that there is no PAH or dust emission and that the MIR SED is consistent with stellar photospheric colors. On the other side of the gap are galaxies with red MIR colors, which means that the SED is dominated by PAHs and warm/hot dust. Galaxies in the gap region would have weak PAH/dust emission compared to starlight. The trend with \ion{H}{1} type is the same as seen in Figure \ref{hcgcol}: galaxies from \ion{H}{1}-rich groups (Type I) generally have ``active'' (red) MIR colors, while galaxies from \ion{H}{1}-poor groups (Type III) generally have ``passive'' (blue) MIR colors. Galaxies from Type II groups fall on both sides of the gap. Interestingly, galaxies within the gap are from either Type I or Type III groups, but not Type II, but given the small number statistics, this may not be significant.

\begin{figure}
  \plotone{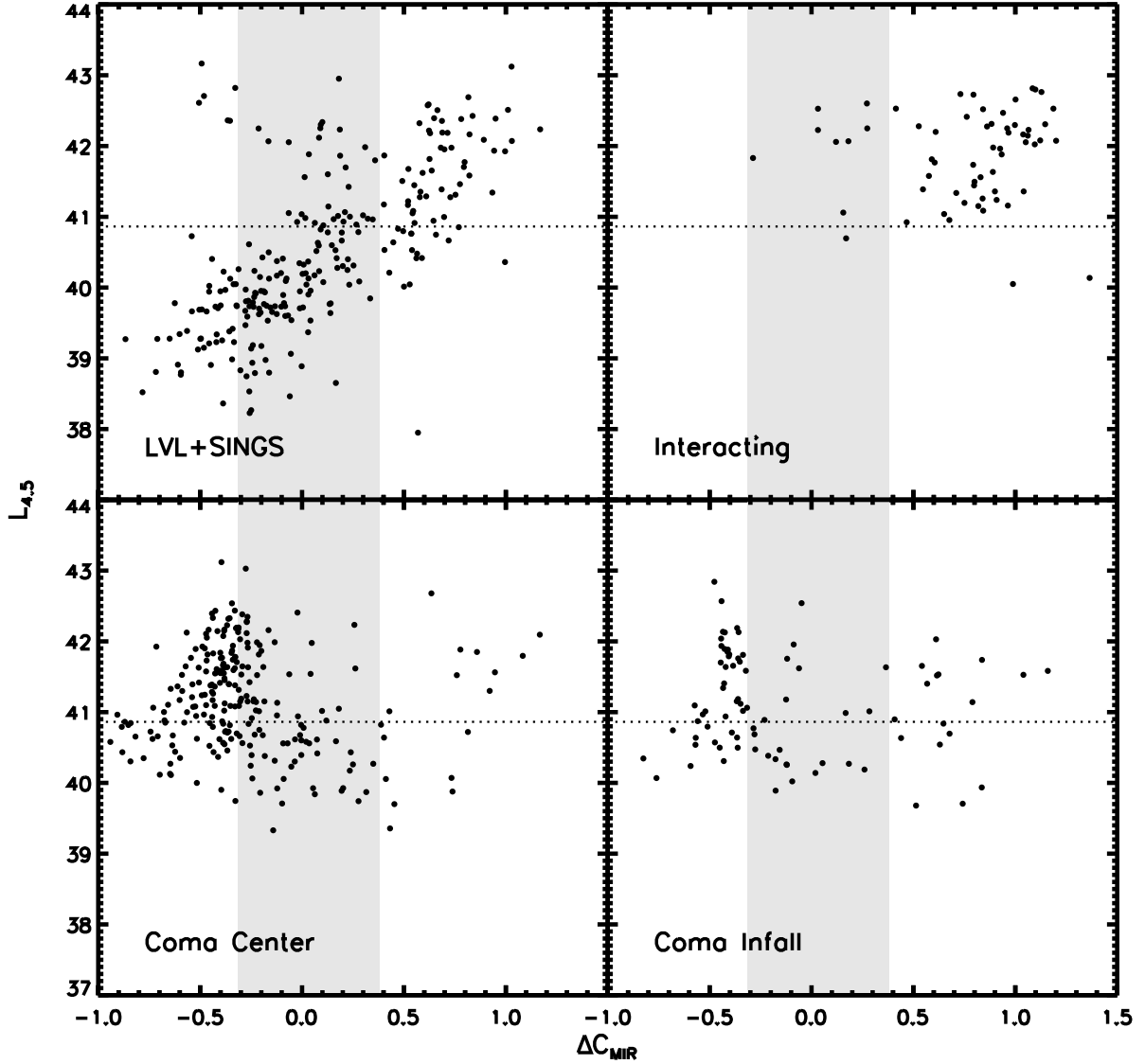}
\caption{CMD for the comparison samples. As in Figure \ref{hcgcmd}, the dotted line indicates the luminosity cut imposed on the samples, and the HCG gap is highlighted by the grey stripe. The criteria used to define each sample are apparent in the shape of the CMDs. LVL+SINGS is the only sample which shows color dependence on luminosity, discussed further in the text.\label{allcmd}}
\end{figure}

The CMDs for the comparison samples are shown in Figure \ref{allcmd}. Like the HCGs, the luminosities of galaxies in the interacting and both Coma samples also do not appear to be related to color over the range of the sample. However, the LVL+SINGS sample does show a correlation between luminosity and color - the brighter galaxies tend to have MIR-redder colors (with the exception of the `tail'). Thus LVL+SINGS has primarily faint, MIR-blue (inactive) galaxies and bright, MIR-red (active) galaxies.

As with the color-color plots, the shape of the CMDs reflects the criteria used to define each sample. For example, the center Coma and infall Coma samples have a significant concentration of galaxies with blue MIR colors. These colors indicate SEDs that fall from 3 to 8 ${\rm \mu m}$, consistent with stellar light arising from the tail of the Raleigh-Jeans distribution, and a general lack of emission from the interstellar medium. The interacting sample, on the other hand, is strongly biased towards actively star forming systems. Therefore its CMD lacks a concentration of massive galaxies with blue MIR colors. Instead, it shows a scattering of galaxies with red MIR colors, indicative of PAH features associated with active star formation or SEDs that rise to longer wavelengths due to the presence of warm dust.

The CMD of the HCGs is a composite -- it contains both a concentration at blue MIR colors, as well as a scattering of systems with red MIR colors, but very few galaxies with intermediate MIR colors. As suggested by the KS tests, the HCG CMD distribution is most similar to that of the sample from the Coma Infall region, and quite unlike that of the other samples.

\section{PROPERTIES OF GALAXIES ALONG MIR COLORSPACE}
\subsection{Morphology}
\begin{figure}
  \plottwo{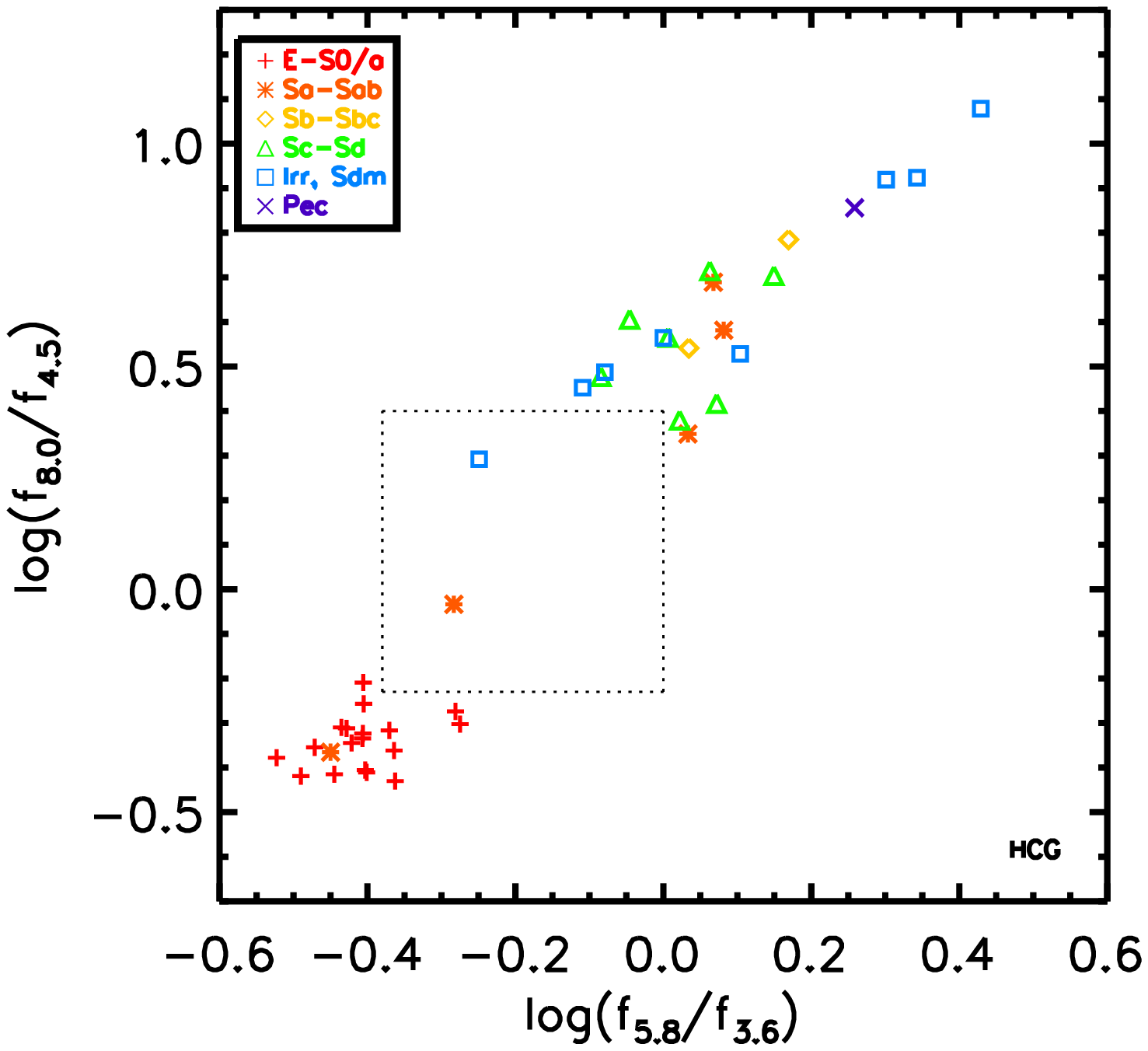}{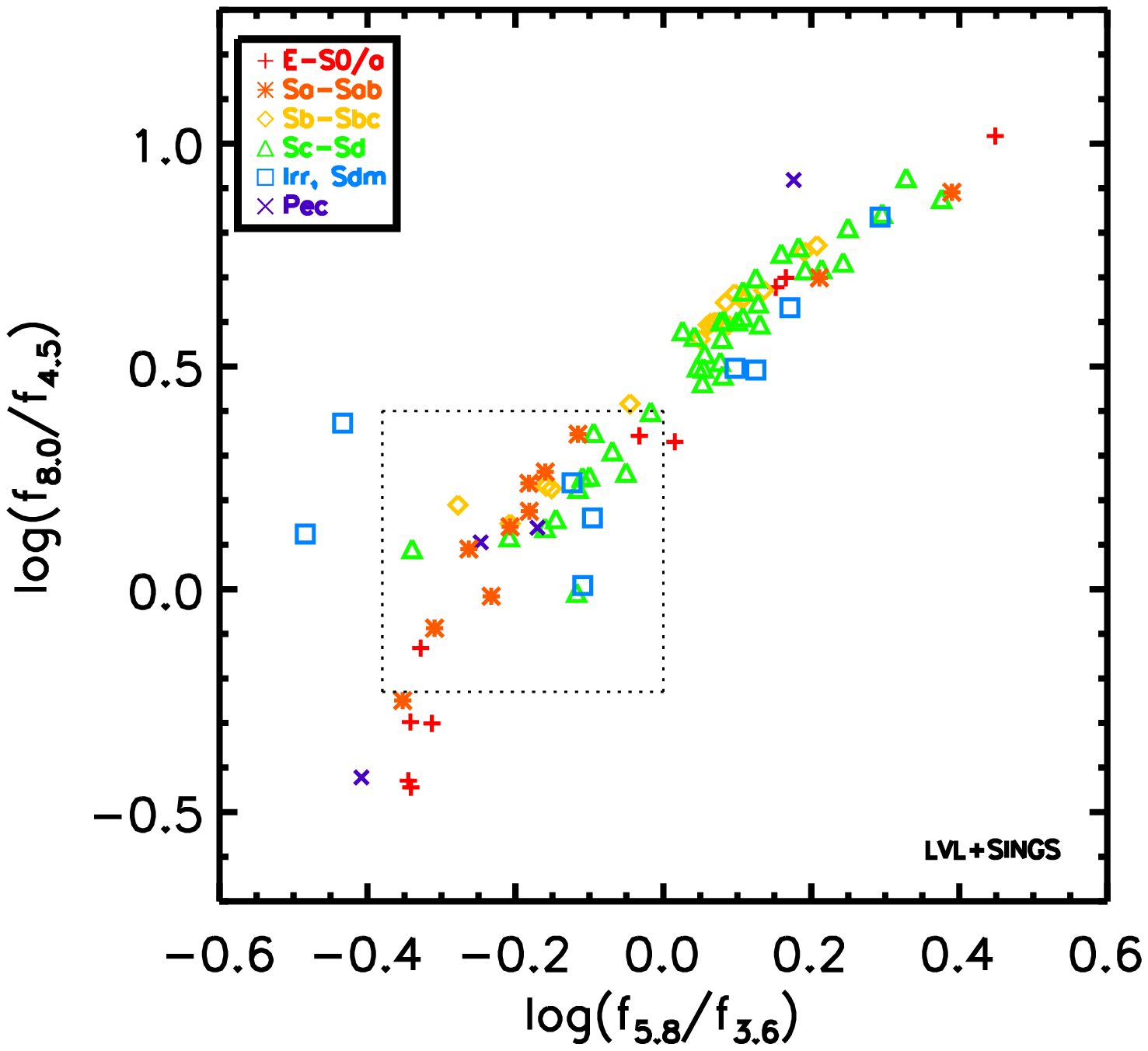}
\caption{Distribution of galaxy morphology in MIR colorspace for {\it left:} HCGs and {\it right:} LVL+SINGS. Note that the distribution of morphologies for the HCGs is as expected, with E/S0 galaxies occupying the lower left and spiral galaxies occupying the upper right. This is not strictly the case for LVL+SINGS.\label{morphcolor}}
\end{figure}
In an attempt to determine the properties of galaxies as a function of MIR colorspace, as well as the physical origin of the gap, we plotted the distribution of morphologies (obtained from \citet{hcgs,sings,lvl}) for the HCG and LVL+SINGS galaxies in colorspace, as shown in Figure \ref{morphcolor}. Interestingly, the LVL+SINGS colors appear to be independent of morphology, while the colors of the HCG sample show a morphological segregation. For the HCG sample, the two galaxies that fall in the gap are Sab and Im. The LVL+SINGS sample shows a variety of morphological types in the gap. We considered the possibility that the lack of correlation was due to the diversity of SINGS galaxies, but when we remove the SINGS sample, the same lack of correspondence is seen in the LVL sample. For the culled SINGS sample of \citet{arp}, we also do not see a strong correlation between Hubble type and Spitzer IRAC colors.

We find this result both surprising and puzzling, especially given previous studies which have found a trend in MIR color with Hubble type \citep{pahre04} and classical work by \citet{kennicutt83} that shows a correlation between H$\alpha$ equivalent width and Hubble type. We have investigated the most extreme examples of this lack of morphological segregation - the MIR red E/S0 galaxies and the Sm/Sdm galaxies which fall in the gap. We found that the blue early type galaxies were all starbursts (see \S\ref{ew}), while the red late type galaxies are low surface brightness galaxies \citep[i.e. NGC 45, 4656, and 5398;][]{monnier03}, have low metallicities \citep[i.e. NGC 55;][]{jackson06}, or are Seyferts which exhibit large flux variations \citep[i.e. NGC 4395;][]{minezaki06}. One possible explanation is that the environment present in compact groups causes morphology to more closely track the activity level of a galaxy than in the field. We hypothesize that this could be due to the presence or absence of neutral gas - if there is gas present, the galaxies will be actively forming stars, which will typically be visible as a disk or irregular galaxy. If most of the gas has been used up, there will be no star formation occurring, and the galaxy will typically be an elliptical or lenticular. In the field, the amount of gas and level of star formation do not appear to be as closely connected, so the galaxy types do not show the segregation seen in the HCG galaxies.

\begin{figure}
  \plottwo{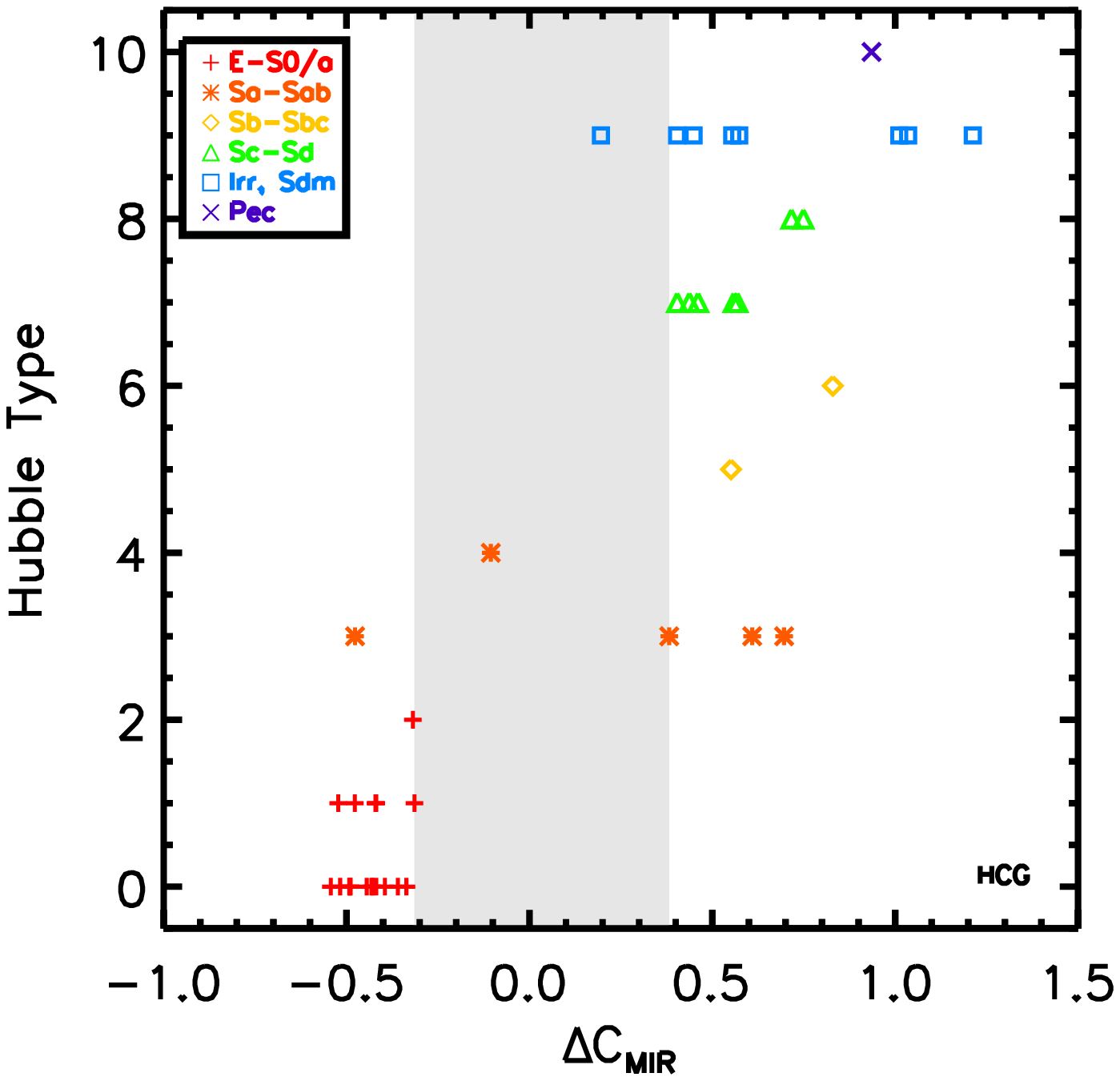}{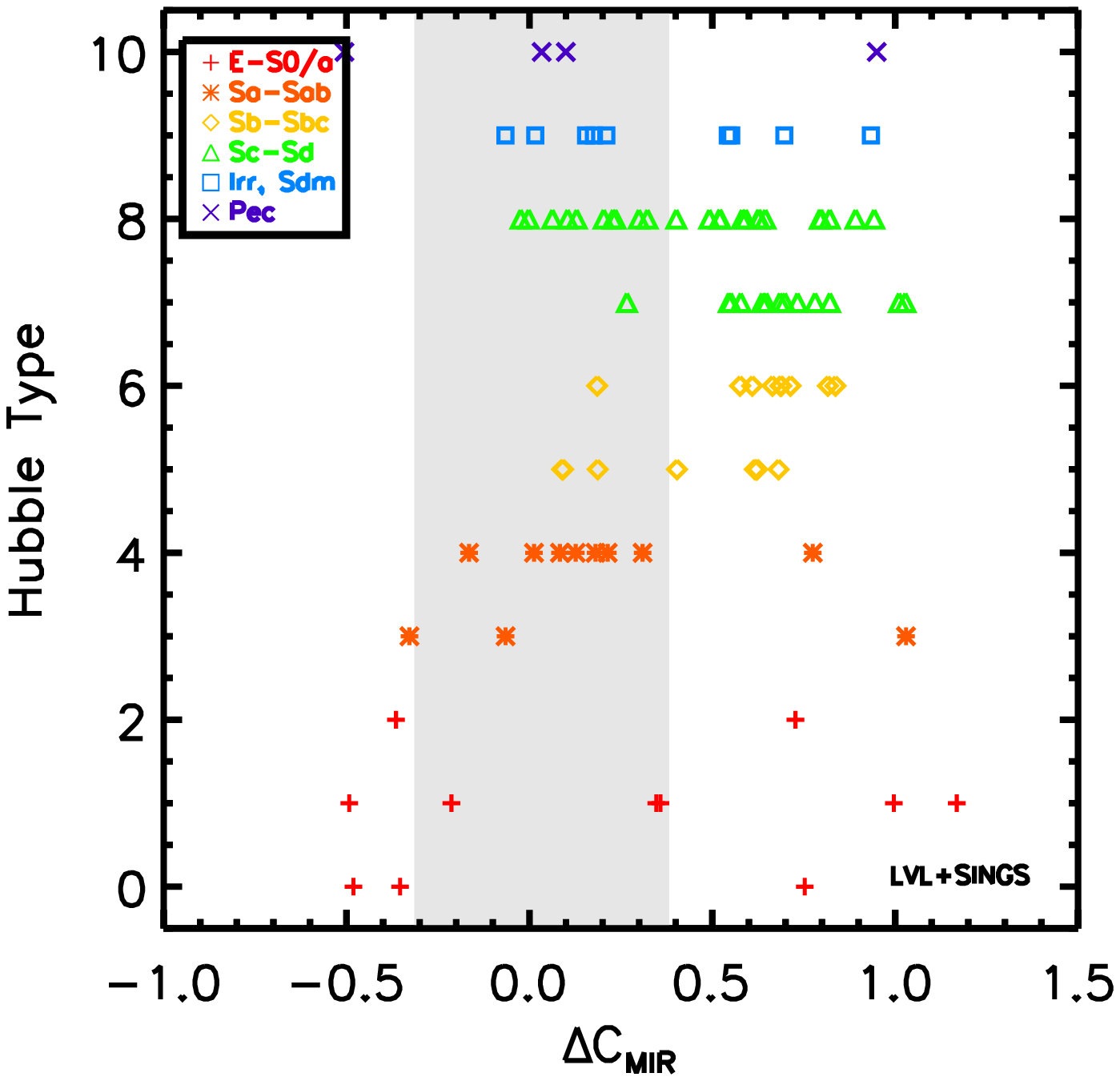}
\caption{Distribution of galaxy morphology in rotated MIR colorspace for {\it left:} HCGs and {\it right:} LVL+SINGS. Note the dearth of morphologies over $2<T<6$ in the HCG sample.\label{morphrotate}}
\end{figure}

Figure \ref{morphrotate} shows Hubble type as a function of rotated MIR color for the HCG galaxies and LVL+SINGS sample. This plot clearly reveals a dearth of galaxies over $2<T<6$ (S0/a-Sbc) for the HCGs, seen previously in Figure \ref{hubblehist}. Of the few galaxies with these Hubble types, they do not preferentially fall in the gap. The LVL+SINGS galaxies do not show a dearth of galaxies between these Hubble types. In addition, almost every Hubble type appears in the gap. This begs the question of whether the existence of the gap is nature or nurture. The selection criteria for the HCGs may have for some reason selected against these morphological types. Alternatively, there might something about the HCG environment that suppresses these Hubble types or alters their SFR so that they do not fall in the gap.

\subsection{EW(H$\alpha$) and Specific Star Formation Rates}\label{ew}
\begin{figure}
  \plottwo{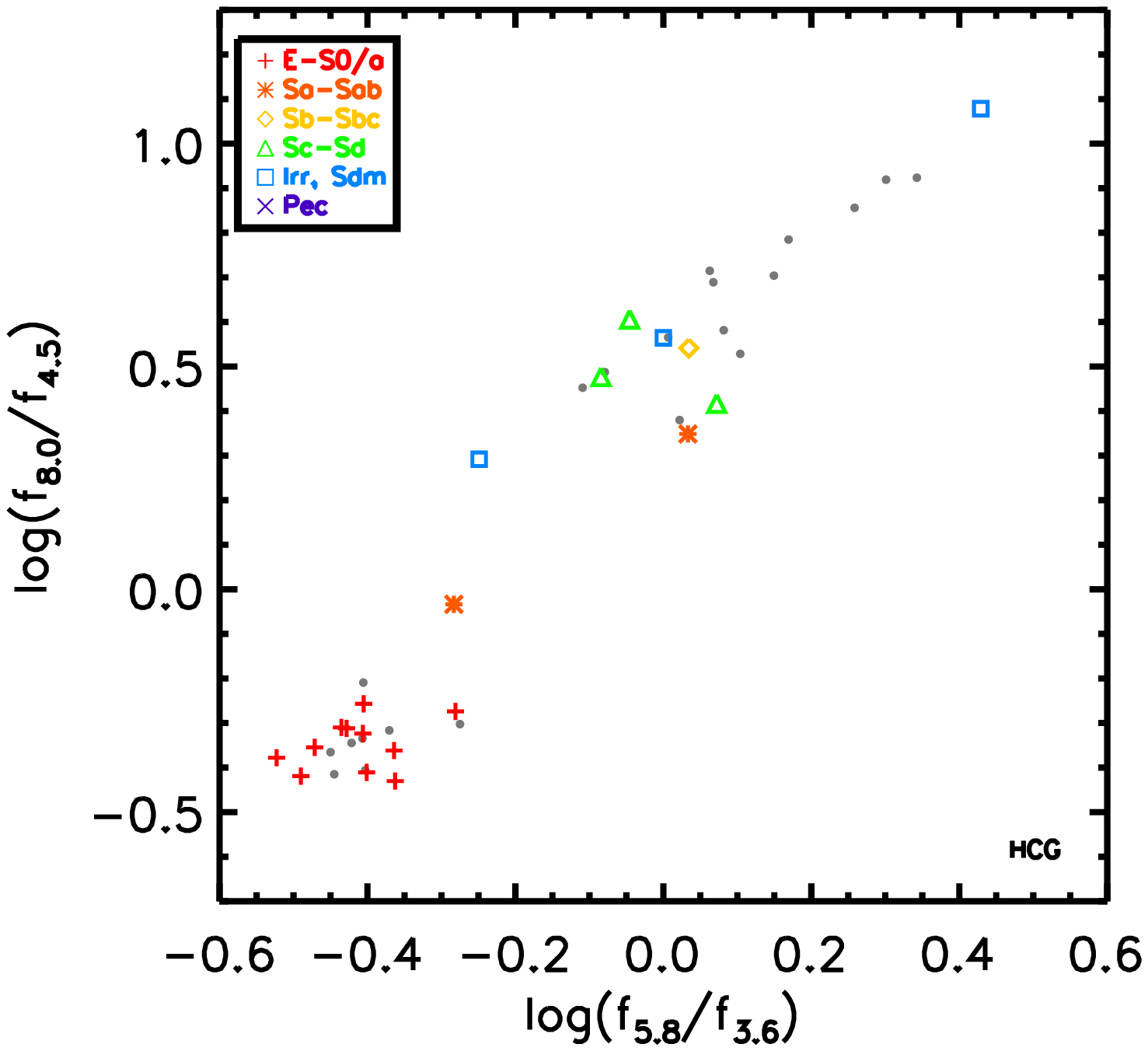}{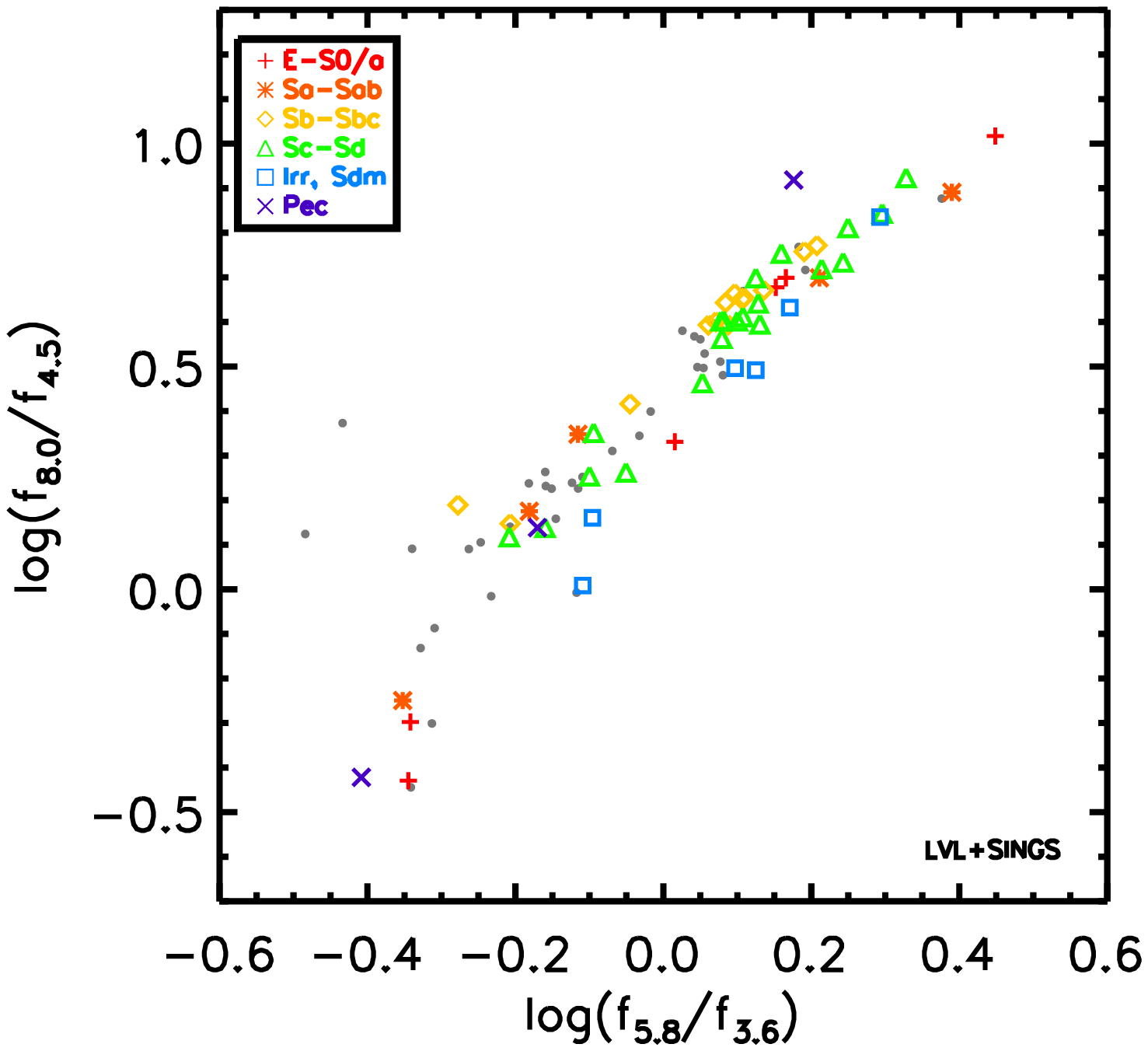}
\caption{IRAC colorspace distribution for {\it left:} HCGs and {\it right:} LVL+SINGS. Galaxies for which spectra were obtained are identified by their morphological type. Galaxies for which spectra were unavailable are shown as grey dots. These plots show that spectra sample the full range of colorspace.\label{spectraid}}
\end{figure}
\begin{figure}
  \plottwo{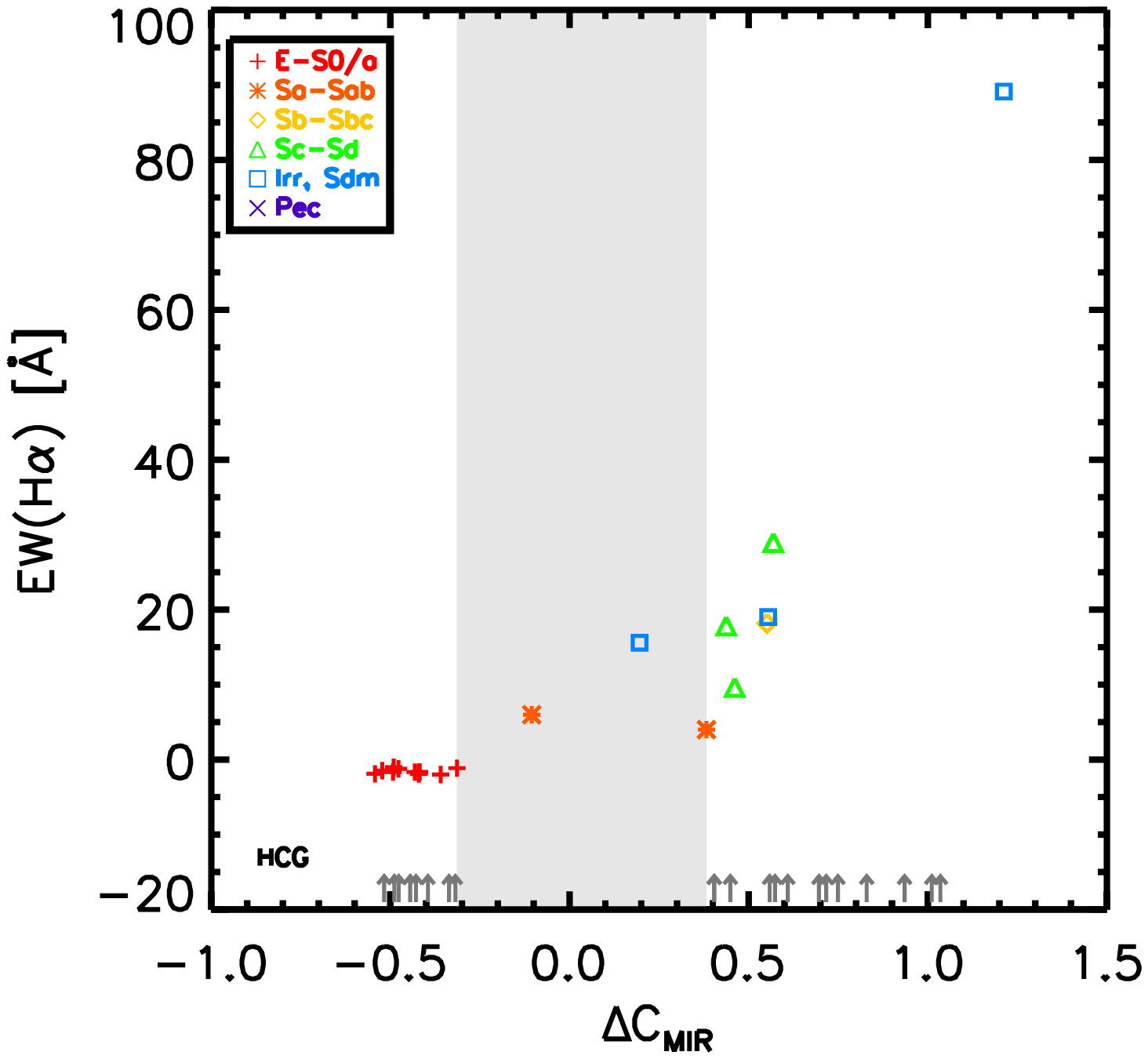}{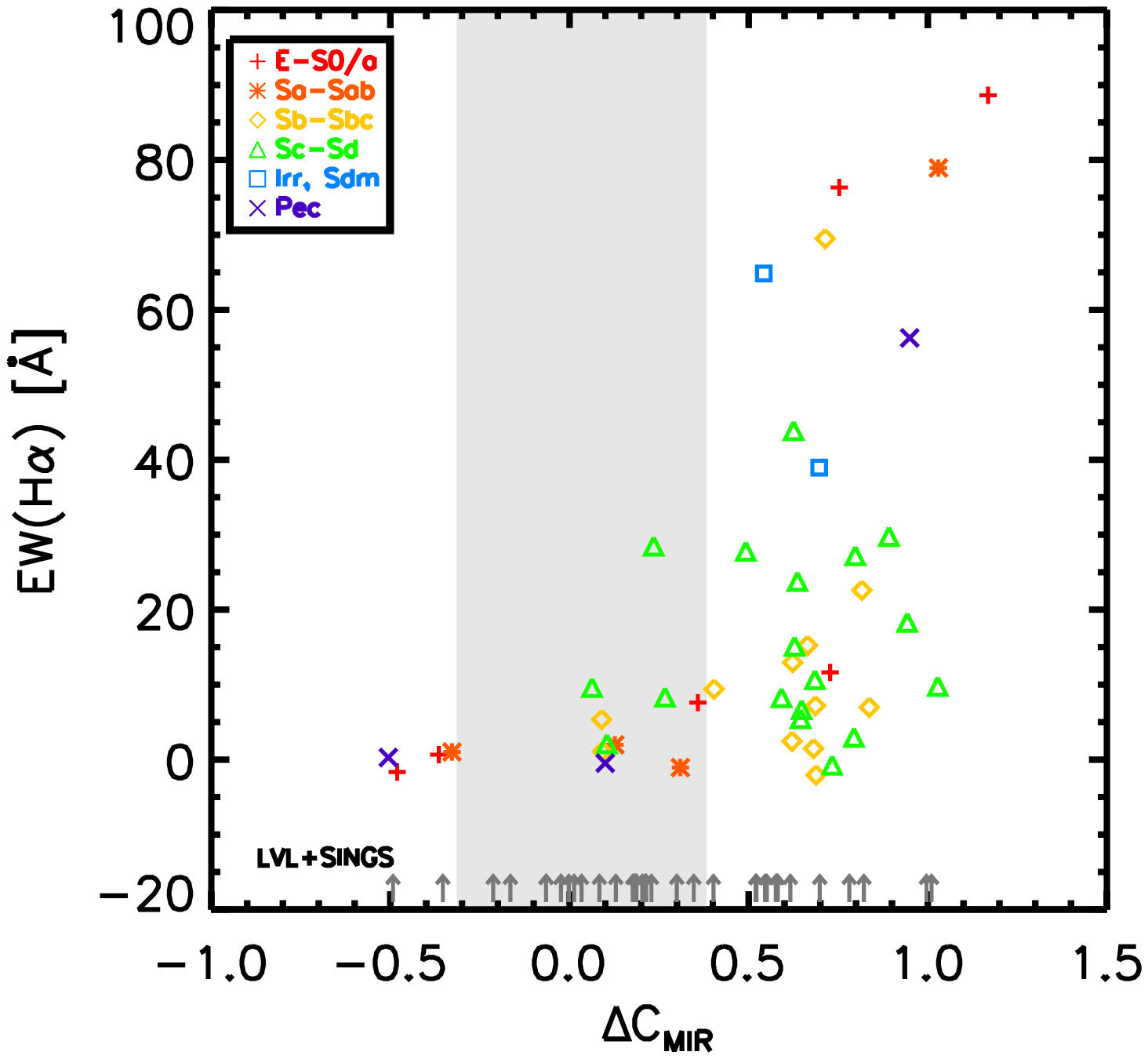}
\caption{$EW\left(H\alpha\right)$ vs $\Delta C_{MIR}$ for {\it left:} HCGs and {\it right:} LVL+SINGS. Galaxies for which spectra were unavailable are plotted as dark grey dots at the bottom. The gap is identified by the grey stripe. In both samples, there is a clear difference between galaxies blueward of the gap and those redward of the gap.\label{ewha}}
\end{figure}

The emission (or absorption) of Balmer lines tracks the recent star formation history of a galaxy through ionized gas, and is complimentary to the MIR data which trace warm dust emission. In order to gain more insight into the properties of star formation in the gap region, here we looked at the $EW\left(H\alpha\right)$ as a function of $\Delta C_{MIR}$. 
For four groups in our HCG sample (HCGs 7, 42, 62, 90), we have obtained Hydra CTIO spectra with $2''$ fibers located on the nuclei. For many of the galaxies in the SINGS sample, optical spectra were obtained with $2.5 \times 2.5''$ slits located on the nuclei \citep{moustakas06}. These data are publicly available at http://irsa.ipac.caltech.edu/data/SPITZER/SINGS/. In addition to these two datasets, many of the galaxies in both the HCG galaxies and in the LVL+SINGS sample have spectra available through SDSS.

For all of these galaxies, we determined the equivalent width of the $H\alpha$ line using Gaussians in the SPLOT task in IRAF. Figure \ref{spectraid} shows the morphologies and locations in colorspace of the galaxies for which we were able to obtain spectra. A plot of $EW\left(H\alpha\right)$ vs $\Delta C_{MIR}$ is shown in Figure \ref{ewha}. In both the HCG sample and the LVL+SINGS sample, all galaxies bluewards of the gap have $EW\left(H\alpha\right)\lesssim 0$, which is fully consistent with MIR colors indicating predominantly stellar emission. Galaxies redward of the gap exhibit a large range of $EW\left(H\alpha\right)$, but there is not necessarily a one-to-one correspondence with MIR emission. This could be explained by several scenarios: 1) dust absorption of $H\alpha$ and general effects of relative geometry; 2) variable amounts of PAH emission; 3) different timescales traced by $H\alpha$ and MIR colors; 4) a coverage difference - the MIR colors are integrated over the entire galaxy, while the $H\alpha$ was measured only in the nuclear region. For a given $\Delta C_{MIR}$, the galaxies in the LVL+SINGS sample cover a larger range in $EW\left(H\alpha\right)$ than the HCG galaxies. This is likely because the galaxies in the LVL+SINGS sample have varying amounts of dust - in fact, the SINGS sample was chosen to be as diverse as possible \citep{kennicutt03}. All galaxies in the HCG sample have $EW\left(H\alpha\right)$ consistent with their Hubble type \citep{nakamura04}. For LVL+SINGS, there are several galaxies which have $EW\left(H\alpha\right)$ inconsistent with their Hubble type - there are peculiar galaxies with $EW\left(H\alpha\right)\sim0$ and E/S0 galaxies with large $EW\left(H\alpha\right)$. In both samples however, the $EW\left(H\alpha\right)$ seems to exhibit an upper envelope, that changes as a function of $\Delta C_{MIR}$.

\begin{figure}
  \plotone{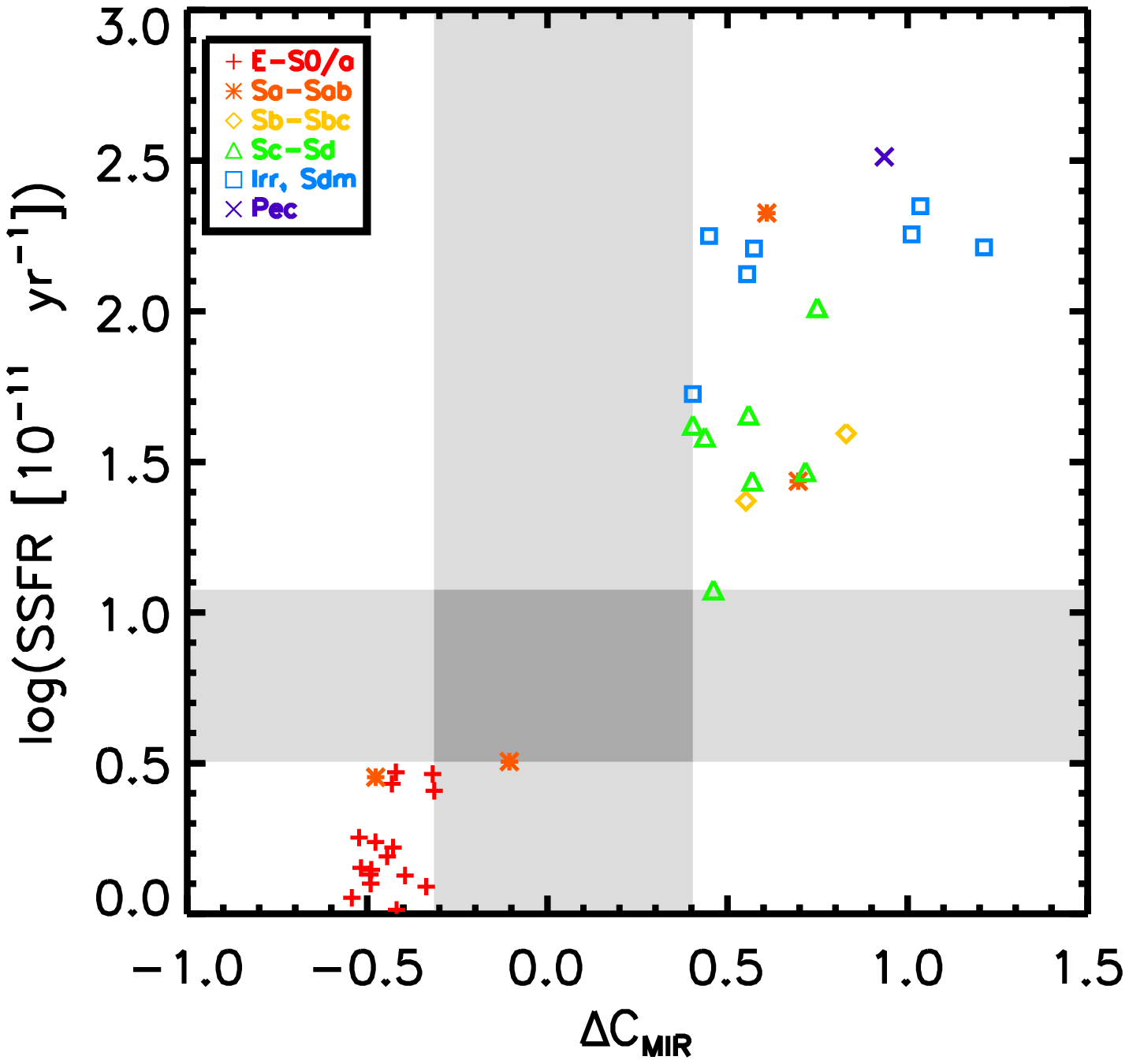}
\caption{SSFR \citep[from][]{tzanavaris10} vs $\Delta C_{MIR}$ for 11 of the 12 HCGs in our sample (HCG 90 was excluded due to insufficient data for determining SSFRs). The vertical grey stripe indicates the gap seen in MIR colorspace, the horizontal grey stripe indicates the gap in SSFR.\label{ssfr}}
\end{figure}

As galaxies in the gap have similar $EW\left(H\alpha\right)$ in both the HCG and LVL+SINGS sample, the properties of the gap galaxies are likely similar. Hence, the HCG sample must be deficient in a certain type of galaxy, rather than having galaxies with fundamentally different properties. Since the gap is between galaxies with colors consistent with stellar photospheric emission and galaxies with relatively high amounts of star formation, we suggest that the gap is likely caused by a deficit of galaxies with moderate specific star formation rates (SSFR). In order to investigate this, we took the SSFRs from \citet{tzanavaris10} and plotted them against MIR color, as shown in Figure \ref{ssfr}. As is clearly visible from the figure, the MIR gap corresponds perfectly with the gap in SSFR. Thus the HCG environment appears to be inhospitable to galaxies with moderate SSFRs ($3.2\times10^{-11} < SSFR < 1.2\times10^{-10}\;{\rm yr^{-1}}$).

\section{DISCUSSION}\label{discussion}
\subsection{The relationship between environment and presence of gap}
Our main result is that the HCG sample is statistically different from the comparison samples in mid-infrared (MIR) colorspace, cumulative distribution functions (CDFs), and rotated color-luminosity diagrams (CMDs), due to the presence of a gap in MIR colorspace. Of the other samples considered here, the HCG sample is most like the infall Coma sample, and unlike the interacting sample, the center Coma sample, or the LVL+SINGS sample. While the HCG sample is most like the infall Coma sample, the lack of galaxies with transitional MIR colors is more pronounced in the HCG sample.

The fact that the HCG sample is most like the infall Coma sample is unsurprising, as other similarities between the compact group environment and the outskirts of clusters have been seen. \citet{lewis02} found suppression of SFRs in clusters out to three virial radii, and determined that star formation suppression depends more on local galaxy density than on overall cluster properties. In addition, studies by \citet{cortese06} of a compact group falling into a cluster suggest that the galaxies in the group have undergone pre-processing due to the local compact group environment.

\subsection{Possible mechanisms for MIR gap}
We have established the statistical significance of the gap in \S\ref{ks}, however the origin and nature of the gap remains an important issue; here we consider three possibilities. The first is that it arises from a fluctuation due to small number statistics; however, the KS tests presented in \S\ref{ks} conclusively demonstrate that the HCG sample is inconsistent with being drawn from a uniform distribution, even taking into account sample size. The second possibility is a subtle selection effect by either us or Hickson, a possibility that can be investigated by including more HCGs in the analysis as well as expanding the sample to include Redshift Survey Compact Groups \citep[RSCGs;][]{barton96}, though it seems unlikely that a selection effect could cause the observed gap. In other words, this would require a bias against galaxies with moderate SSFR, but not low or high SSFR. The third possibility is that the paucity of morphological types could correspond to the MIR gap. The morphology histogram (Figure \ref{hubblehist}) shows a dearth of galaxies between $2 \ge T \ge 6$. For the HCG sample it is impossible to determine what is {\it not} in the gap, but we can look at gap galaxies in LVL+SINGS for insight. Since LVL+SINGS shows galaxies of all types in the gap, it is not consistent with morphological type being a predominant factor in creating the gap. However, it is possible that HCGs select against galaxies that would fall in the gap. This leads us to the possiblity that the gap is due to a deficit of galaxies exhibiting moderate specific star formation rates due to environmental effects of the CG environment.

We conclude that the distribution in colorspace, and in particular the occurence of a gap in the MIR color distribution most likely reflects different levels of specific star formation, with galaxies ranging from active (red MIR colors) to passive (blue MIR colors). This is supported by the gap seen in the SSFRs for these HCG galaxies, independently determined using UV, 24$\mu m$, and K-band data \citep{tzanavaris10}. The distribution in colorspace suggests that galaxies in the HCG environment spend little time in the evolutionary state that yields MIR colors in the gap region. The fact that the gap is seen in both the HCG and Coma infall samples, but is more pronounced in the HCG sample indicates that this effect is similar to, but more dramatic than, the environment in the outskirts of clusters. The tidal fields present in HCGs can easily funnel gas to the inner parts of galaxies, inducing high levels of star formation. Once this gas is used up or heated to high temperatures, the galaxies can no longer form stars, and quickly become quiescent. Thus galaxies in HCGs either experience a profusion of star formation or do not have the gas necessary to form stars, and do not exhibit moderate levels of star formation.

\subsection{Relationship to optical CMDs}\label{optcmds}
Insight into the shape of MIR CMDs may be gained by analogy to optical CMDs. In the latter, several prominent features have been identified - an optical ``red sequence'' of galaxies with a relatively narrow distribution of red optical colors, extending to the brightest magnitudes; an optical ``blue cloud'' of generally fainter galaxies with bluer optical colors \citep[e.g.][]{strateva01,hogg04}; and a deficit of galaxies with intermediate optical colors (the so-called ``green valley''; \citet{hogg04}). Optical red sequence galaxies are dominated by E/S0 galaxies, while optical blue cloud galaxies are dominated by disk and irregular systems. This optical CMD shape is apparent in both dense and sparse environments, although dense environments have a more prominent red sequence, while less dense environments have a more pronounced blue cloud.

The optical CMD shape is often interpreted in terms of evolutionary processes, in which optical blue cloud galaxies might evolve onto the red sequence through some combination of wet and dry merging, star formation, star formation ``quenching'', and aging \citep[e.g.][and references therein]{faber07}. A key attribute of such models is that galaxies must move rapidly between the optical blue and red phases in order to reproduce the red/blue dichotomy \citep[e.g.][]{dekel06}.  The optical ``green valley'' region therefore includes these evolutionary transitional systems. However, the optical red sequence is contaminated by dusty star-forming galaxies. This is a significant advantage of the MIR over optical.

The interpretation of the MIR CMDs is expected to be similar to that of the optical CMDs, however discussing MIR and optical CMDs simultaneously can be confusing because of the different connotations of ``blue'' and ``red''. In optical colors blue light typically reflects a young population and the optical red sequence picks out both ``red and dead'' galaxies (i.e. galaxies whose colors are dominated by evolved stars {\it and} have little or no active star formation \citep[e.g.][]{vandokkum05,bell04}) and galaxies obscured by dust, which may be active. On the other hand, MIR colors reflect the properties of the dust, so blue MIR galaxies select for systems with little dust contribution, which are dominated by stellar photospheric emission (whether from young or old stars), with little or no active star formation. Blue MIR colors therefore select both ``optically red and dead'' and ``optically blue and dying'' galaxies. Red MIR galaxies are active, with current star formation heating the dust and exciting the PAHs, so they are most closely related to the optical blue cloud. Additionally, the optical green valley is present in all environments \citep{hogg04}, while we have only found evidence for the MIR gap in dense environments that still contain neutral gas.

\subsection{The role of gas in CG evolution}
The presence of neutral gas may be a key factor contributing to the gap in MIR colorspace. Without fuel for star formation, the MIR colors would be dominated by stellar emission. There are currently two theories on \ion{H}{1} distribution within HCGs: the first is that groups whose \ion{H}{1} is contained entirely within the member galaxies are less evolved or younger in stellar population than those groups with \ion{H}{1} that is distributed within and between member galaxies \citep[cf.][]{williams90, verdes01}; the second is that the distribution of \ion{H}{1} determines how HCGs evolve, i.e. HCGs whose \ion{H}{1} is contained entirely within the galaxies evolve differently than HCGs whose \ion{H}{1} is distributed throughout the group \citep{iraklis10}. Thus the \ion{H}{1} properties of a group are a crucial component related to a galaxy's MIR color. HCG galaxies from \ion{H}{1}-rich groups (Type I) primarily lie redward of the gap, while galaxies from \ion{H}{1}-poor groups (Type III) primarily lie blueward. Galaxies from Type II groups lie both redward and blueward of the gap, while avoiding it. Curiously, the few galaxies within the gap are from either Type I or Type III groups. Further investigation into this trend will necessitate interferometric \ion{H}{1} observations of compact groups, in order to determine the \ion{H}{1} deficiency (and thus type) of individual galaxies within compact groups. \ion{H}{1} imaging will reveal whether ``rogue'' galaxies (e.g. MIR-red galaxies from Type III groups) have a different individual deficiency from their group as a whole. Interferometric observations will also allow us to determine where the \ion{H}{1} is located - whether it is confined to the member galaxies, or distributed throughout the group. \ion{H}{1} distribution could be another clue in compact group evolution.

One estimate of the timescale of the proposed rapid evolution in SSFR is the time it takes a galaxy to use up the available gas and go from starburst to poststarburst. Thus this could occur on timescales as short as a million years \citep{gao99}, but will clearly be highly dependent on environment. A crude upper limit (assuming 100\% efficiency and a constant SFR), using the \ion{H}{1} mass of the group divided by the sum of the SFRs in each member galaxy yields gas depletion timescales ranging from 0.4 Gyr to 4 Gyr.

The presence of the gap in the MIR colorspace distribution of the HCGs combined with the fact that the gap is not present in less dense environments indicates that local environment significantly influences galaxy properties. In order to understand the processes that affect galaxy evolution, we need to understand how gas is processed in the interstellar medium and intragroup medium distinct from the field and cluster environments. Compact groups are clearly an important part of understanding galaxy evolution and cluster assembly, especially considering the similarities between the distribution in colorspace of the HCGs and the Coma infall region.

\acknowledgments
K.E.J. gratefully acknowledges support for this paper provided by NSF through CAREER award 0548103 and the David and Lucile Packard Foundation through a Packard Fellowship. S.C.G. thanks the National Science and Engineering Research Council of Canada for support. For helpful discussions on statistical tests, L.M.W. thanks statistics professor Tao Huang. We also thank the anonymous referee for their constructive comments. This research has made use of the NASA/IPAC Extragalactic Database (NED) which is operated by the Jet Propulsion Laboratory, California Institute of Technology, under contract with the National Aeronautics and Space Administration.

\begin{multicols}{2}

\end{multicols}

\clearpage
\end{document}